\documentclass[11pt,notitlepage,a4paper]{article}
\pdfoutput=1
\usepackage[a4paper,margin=1in]{geometry}
\usepackage{amsmath,amssymb,amscd,amsfonts}
\usepackage{mathtools}
\numberwithin{equation}{section}
\usepackage{url}
\usepackage[shortlabels]{enumitem}
\usepackage{tikz} 
\usetikzlibrary{decorations}
\usetikzlibrary{arrows,decorations.markings}         
\usepackage[hidelinks]{hyperref}
\usepackage[open,
  openlevel=2,
  atend,
  numbered]{bookmark}
\usepackage{graphicx}
\usepackage[final]{pdfpages}
\usepackage{threeparttable}
\usepackage[small,sc]{caption}
\usepackage{float}
\usepackage{stackengine}
\usepackage{scalerel}
\usepackage[bf,small,center]{titlesec}
\usepackage{tocloft}

\newlength{\mylen}	
\urldef\footurlb\url{cocalc.com/dfriedan/DM/SM}
\urldef\footurla\url{physics.rutgers.edu/~friedan}
\usepackage{sidecap}
\usepackage{cite}
\def\eq{\begin{equation}}
\def\en{\end{equation}}
\def\eqg{\eq\begin{gathered}}
\def\eng{\end{gathered}\en}
\def\eqa{\eq\begin{aligned}}
\def\ena{\end{aligned}\en}
\def\Reals{\mathbb{R}}

\def\expval#1{\langle \, #1 \,\rangle}

\def\Complexes{\mathbb{C}}

\def\cM{\mathcal{M}}

\def\bK{\mathbf{K}}
\def\bM{\mathbf{M}}

\stackMath
\def\dyhat{-0.2ex}
\newcommand\myhat[1]{\ThisStyle{%
              \stackon[\dyhat]{\SavedStyle#1}
                              {\SavedStyle\hat{\phantom{#1}}}}}
\def\that{\kern0.1em\myhat{\kern-0.1em t}}

\def\texttilde{\raise-0.7ex\hbox{\!\texttt{\char`\~}}}
\titlespacing{\section}{3pc}{1.75pc}{0.8pc}
\titleformat{\section}
  {\centering\normalsize\bfseries} 
  {\large\thesection}
  {0.5em} 
  {\large}
\titleformat{\subsection}
  {\centering\normalsize\bfseries} 
  {\thesubsection}
  {0.5em} 
  {}
\titleformat{\subsubsection}
  {\normalsize\bfseries} 
  {\thesubsubsection}
  {0.5em} 
  {}
\usepackage{diagbox}
\usepackage{amsthm}
\def\tM{\tilde M}
\def\tN{\tilde N}
\def\tA{\tilde A}

\def\conn{\mathrm{c}}

\def\hcM{\widehat\cM}

\def\cN{\mathcal{N}}
\def\tNc{\tN_{\conn}}

\def\bn{\mathbf{n}}
\def\bny{\bn\bracky}

\def\bm{\mathbf{m}}
\def\Diff{\mathrm{Diff}}
\def\MCG{\mathrm{MCG}}

\def\dmo{$(d{-}1)$}

\def\NN{\mathbb{N}}
\def\expval#1{\left\langle#1\right\rangle}

\def\Sym{\otimes_{s}}

\newtheorem*{Theorem*}{Theorem}

\def\cS{\mathcal{S}}

\def\tcS{\tilde\cS}
\def\tO{\tilde{O}}

\def\iprod{\mathrel{\raisebox{2pt}{$\lrcorner$}}}

\def\bA{\mathbf{A}}

\def\bPhi{\boldsymbol\varPhi}
\def\bphi{\boldsymbol\varphi}

\def\bcS{\boldsymbol{\cS}}

\def\cSd{\bcS}

\def\bq{\mathbf{q}}
\def\bn{\mathbf{n}}
\def\bS{\mathbf{S}}

\def\textprod{\mathop{\textstyle \prod}\limits}
\def\textsum{\mathop{\textstyle \sum}\limits}

\def\Ox{O\brackx}

\def\bQ{\mathbf{Q}}
\def\phic{\phi_{\conn}}

\def\Mx{M\brackx}

\def\bw{\mathbf{w}}
\def\Ny{N\bracky}

\def\obs{\text{obs}}

\def\bJ{\mathbf{J}}
\def\bU{\mathbf{U}}

\def\couple{\mathop\Bumpeq}

\def\btau{\boldsymbol\tau}
\def\btcS{\boldsymbol{\tcS}}
\def\Sym{\mathrm{Sym}}
\def\hcM{\hat\cM}

\def\meas{\mathrm{meas}}

\def\cSy{\cS\bracky}
\def\bcSx{\bcS\brackx}
\def\bZ{\mathbf{Z}}
\def\bW{\mathbf{W}}

\def\sensor{\mathrm{sensor}}

\def\cSy{\cS\bracky}
\def\Ny{N\bracky}
\def\cX{X}
\def\cY{Y}

\def\wred{w_{\mathrm{red}}}

\def\bM{\mathbf{M}}
\def\bz{\mathbf{z}}

\def\bv{\mathbf{v}}

\def\couple{\boldsymbol{\ltimes}}

\def\sslbrack{{\scriptscriptstyle\lbrack}}
\def\ssrbrack{{\scriptscriptstyle\rbrack}}
\def\sbrack#1{\sslbrack #1 \ssrbrack}
\def\slbrack{{\scriptstyle\lbrack}}
\def\srbrack{{\scriptstyle\rbrack}}
\def\brack#1{\slbrack #1 \srbrack}
\def\brackx{\brack{x}}
\def\bracky{\brack{y}}
\def\GammaX{\Gamma_{X}}
\def\GammaY{\Gamma_{Y}}
\def\Gammasensor{\Gamma_{\sensor}}

\def\bp{\mathbf{p}}
\def\bu{\mathbf{u}}
\def\hw{\hat w}
\begin{document}
%
%

\begin{center}
{\Large Global structure of euclidean quantum gravity}
\vskip5ex
{\large Daniel Friedan}
\vskip2ex
{\small\it
New High Energy Theory Center
and Department of Physics and Astronomy\\
Rutgers, The State University of New Jersey\\
Piscataway, New Jersey 08854-8019 USA
\vskip0.5ex
\href{mailto:dfriedan@gmail.com}{dfriedan@gmail.com}
\qquad
%
\href{https://physics.rutgers.edu/\textasciitilde friedan/}
{physics.rutgers.edu/{$\scriptstyle \sim$}friedan}
}
\vskip2ex
\today
\end{center}
%
%
\begin{center}
\vskip3ex
{\sc Abstract}
\vskip3ex
\parbox{0.96\linewidth}{
\hspace*{1.5em}
%
Euclidean quantum gravity (EQG) separates into a local theory
and a global theory.
The local theory operates in every compact $d$-manifold with boundary
to produce a state on the boundary.
The global theory then sums these boundary states
over the diffeomorphism classes of $d$-manifolds with boundary
to make the Hartle-Hawking state.
Global EQG is formulated here as classical statistical physics.
The Hartle-Hawking state is the probability measure of 
a mathematically natural classical statistical system,
analogous to the functional measure of euclidean quantum field theory.
General principles of global EQG determine
the numerical weights $w(M)$ in the sum over diffeomorphism 
classes $M$.
}
\end{center}

\vspace*{-5ex}

%
%
%
\begin{center}
\tableofcontents
\end{center}
%
%
\addtocontents{toc}{\vspace*{-0.5ex}}
\section{Introduction}

Euclidean quantum gravity (EQG) originated
with 
Hartle and Hawking's proposal \cite{Hartle:1983ai}
to construct a ``quantum gravity state'' 
on closed 3-manifolds
by summing over all compact 4-manifolds with boundary.
They proposed the resulting state as
initial quantum state of the universe,
the ``wave function of the universe''.
A different use is proposed here.
The Hartle-Hawking state is 
the probability measure
of a classical statistical system.

The Hartle-Hawking state is constructed in two steps.
A diffeomorphism invariant functional integral on each 4-manifold with boundary
produces a state on the boundary --- the \emph{amplitude}.
This is the \emph{local} EQG.
Then the amplitudes are summed over all diffeomorphism classes 
of $4$-manifolds with boundary.
The second step is the \emph{global} EQG.

Global EQG is formulated here as classical statistical physics ---
\emph{statistical} EQG
--  in arbitrary space-time dimension $d$.
A local EQG is assumed to be given and to
satisfy certain basic axioms;
otherwise the local theory is unspecified.
The global structure is mathematically natural,
following the general covariance principle of physics.

A classical statistical system consists of (1) a set of objects, 
(2) a space of classical states of the objects,
and (3) a probability measure on the space of states.
Observables are functions on the space of states.
The expectation value of an observable is the integral
with respect to the probability measure.
Euclidean quantum field theory is an example.
The objects are labelled by the points $x$ in 
space-time.
The classical states of the object at $x$
are the possible values $\phi(x)$ of the classical fields at $x$.
The space of classical states of the system --- the space of 
classical fields ---
is the direct product over all the objects $x$ of the states at $x$.
The probability measure is the normalized functional measure on the 
classical fields.
Observables are functions on the space of states --- functions on the space of fields. 
Correlation functions of observables
are their joint expectation values in the probability measure on
fields.

Statistical EQG has two layers: a microscopic system and a 
macroscopic system (the observers).
The microscopic system consists of a set $\cY$ of microscopic objects 
$y$, 
a vector space $\GammaY$ of microscopic states $\phi$, and 
a probability measure $D\phi$ on $\GammaY$.
A microscopic state $\phi$ is a state $\phi\bracky$ for each object 
$y\in Y$.
The macroscopic system consists of a set $X$ of macroscopic objects
$x$
(the observers),
a vector space $\GammaX$ of observer states $\bphi$,
and a probability measure $D\bphi$ on the observer states.
A macroscopic state $\bphi$ is a state $\bphi\brackx$ for each observer
$x\in X$.

There is a natural polynomial
function $T\colon \GammaY\rightarrow \GammaX$
from the vector space of microscopic states to the vector space of 
observer states.
The microscopic state $\phi$  determines the observer state
$\bphi = T(\phi)$.
The microscopic probability measure $D\phi$ on 
$\GammaY$ is transformed by $T$ 
into the macroscopic probability distribution $D\bphi=T_{*}D\phi$ on $\GammaX$.

The local EQG provides a vector space of states on each closed 
\dmo-manifold --- each \dmo-manifold without boundary.
Diffeomorphisms of the \dmo-manifold act on the states.
The \emph{globally invariant} states
are the states of the \dmo-manifold that are invariant under the 
entire
diffeomorphism group of the \dmo-manifold.
The globally invariant states of a \dmo-manifold
are a property of the diffeomorphism 
class of the \dmo-manifold.

Every diffeomorphism of a $d$-manifold with boundary takes the boundary 
to itself, acting as a diffeomorphism of the boundary.
The  \emph{locally invariant} states of the $d$-manifold are the 
states on its boundary
that are invariant under 
all diffeomorphisms of the $d$-manifold.
This is a subgroup of the entire group of diffeomorphisms of the 
boundary.
The locally invariant states of a $d$-manifold are a property of the diffeomorphism 
class of the $d$-manifold.

The set of microscopic objects is the set $Y=\{\Ny\}$
of diffeomorphism classes of \emph{connected} closed \dmo-manifolds 
$\Ny$.
The states $\phi\bracky$ of $y$ are the globally invariant states
of the local EQG
on $\Ny$.
The vector space $\GammaY$ of microscopic states
is the direct product of the globally invariant state
spaces of all the $y\in Y$.
A microscopic state $\phi\in\GammaY$ is a globally invariant state $\phi\bracky$ 
for each microscopic object $y\in Y$.

The set of macroscopic objects (observers) is the set $X=\{\Mx\}$ 
of diffeomorphism classes of \emph{connected} $d$-manifolds with 
\emph{non-empty} boundary.
The states $\bphi\brackx$ of
observer $x$ are the locally invariant states
of $\Mx$.
The vector space $\GammaX$ of macroscopic states
is the direct product of the locally invariant state
spaces of all the observers $x\in X$.
A macroscopic state $\bphi\in\GammaX$ is a locally invariant state 
$\bphi\brackx$ 
for each observer $x\in X$.

For every observer $x\in X$
the globally invariant states on the boundary of $x$ are a subspace of the 
locally invariant states.
Each microscopic  state $\phi\in \GammaY$ 
thus determines a macroscopic state $\bphi = T(\phi) \in \GammaX$.
The boundary of $x$ is a finite disjoint union of connected 
components, so $T(\phi)\brackx$ is a polynomial in the $\phi\bracky$.
The transformation $T\colon \GammaY \rightarrow \GammaX$
is a polynomial function.

The local EQG also provides a particular state --- the \emph{amplitude} ---
on the boundary of each compact $d$-manifold with boundary.
The amplitude is a locally invariant state --- invariant under all 
the diffeomorphisms of the $d$-manifold.

The Hartle-Hawking state on a general closed \dmo-manifold
is a weighted sum of the amplitudes over the ensemble of diffeomorphism
classes of $d$-manifolds that have the \dmo-manifold as boundary.
The collection of Hartle-Hawking states on all the closed \dmo-manifolds
will be seen to define a probablity measure $D\phi$ on $\GammaY$.
This is the probability measure of the microscopic statistical system.

\vskip2ex
\noindent
\hspace{2em}{\bf microscopic system}
\begin{itemize}[itemsep=1ex,topsep=1.5ex,leftmargin=5em]
\item microscopic objects\\[0.5ex]
\hspace*{2em} $Y=\left\{\Ny\right\}=\left\{\text{diffeomorphism 
classes of connected closed 
\dmo-manifolds}\right\}$
\item vector space of states of each object\\[0.5ex]
\hspace*{2em} $\cSy = \left\{\text{globally invariant states on } 
\Ny\right\}$
\item vector space of microscopic states\\[0.5ex]
\hspace*{2em} $\GammaY = $ the direct product 
of the $\cSy$\\[0.5ex]
\hspace*{2em}
$\phi\in \GammaY$ is a globally invariant state $\phi\bracky\in\cSy$
for each $y\in Y$.
\item probability measure $D\phi$ on $\GammaY$.
\end{itemize}

\vskip2ex

\noindent
\hspace{2em}{\bf macroscopic system}
\begin{itemize}[itemsep=1ex,topsep=1.5ex,leftmargin=5em]
\item macroscopic objects (observers)\\[0.5ex]
\hspace*{2em} $X=\left\{\Mx\right\}=\left\{\text{diffeomorphism classes of connected
$d$-manifolds with}\right.$\\
\hspace*{10em}  $\left.\text{non-empty boundary}\right\}$
\item vector space of states of each observer\\[0.5ex]
\hspace*{2em} $\bcSx = \left\{\text{the locally invariant states on } 
\Mx \right\}$
\item vector space of observer states\\[0.5ex]
\hspace*{2em} $\GammaX = $ the direct product 
of the $\bcSx$\\[0.5ex]
\hspace*{2em}
$\bphi\in \GammaX$ is a globally invariant state $\bphi\brackx\in\bcSx$
for each $x\in X$.
\item probability measure $D\bphi=T_{*}D\phi$ on $\GammaX$.
\end{itemize}

Measurements are made by a collection of macroscopic observers 
coupled to the microscopic system.
An \emph{experiment} is the disjoint union of a collection of 
observers, i.e.~a diffeomorphism class of $d$-manifolds with no 
closed components.
The experiment is connected at its boundary to the ensemble of 
diffeomorphism classes of $d$-manifolds.
The Hartle-Hawking state is the boundary condition on the experiment
--- the probabilistic state of the experiment.

The Hartle-Hawking state must be globally invariant in order to 
provide a locally invariant boundary condition to every observer.
This 
is the first constraint on the weightings in the sum over 
diffeomorphism classes of $d$-manifolds.
The sum over the ensemble 
must combine the locally invariant amplitudes
into a globally invariant Hartle-Hawking state.

Each closed \dmo-manifold is a disjoint 
union of connected closed \dmo-manifolds, i.e.~a collection of microscopic 
objects $\Ny$ with multiplicities.
The diffeomorphism group of the boundary permutes identical $\Ny$ in the boundary,
so the globally invariant Hartle-Hawking state is an element in the symmetric algebra 
$\Sym(\GammaY)$,
the symmetric tensor products of the microscopic states $\GammaY$.
In quantum mechanics, $\Sym(\GammaY)$ is the Fock space on 
$\GammaY$.  An inner product on $\GammaY$ makes $\Sym(\GammaY)$ a 
Hilbert space.  But there is a more natural
interpretation of 
$\Sym(\GammaY)$ that does not assume an inner product on $\GammaY$.
The dual space $\GammaY^{*}$ is the space of linear functions on 
$\GammaY$.  $\Sym(\GammaY^{*})$ is the space 
of polynomial functions on $\GammaY$.  The symmetric algebra 
$\Sym(\GammaY)$ is the space of linear functions on 
$\Sym(\GammaY^{*})$ --- the space of linear functions on the 
polynomial functions on $\GammaY$.
So $\Sym(\GammaY)$ is naturally interpreted
as a space  of measures on $\GammaY$.
Multiplication in the symmetric algebra is the convolution of measures
on $\GammaY$.
This is analogous to the structure of statistical field theory.
The functional measure of statistical field theory
is a linear function on the symmetric algebra 
generated by the linear observables on the vector space of classical fields.

$\Sym(\GammaY)$ is formally the same as
the Fock space of ``baby universes'' \cite{Banks:1988je} except
that the elements of $\Sym(\GammaY)$ are not square integrable states in a 
Hilbert space but rather measures on the vector space $\GammaY$.
(And there is no parent universe.
Baby universes bubble in and out of a macroscopic 
parent universe.  In statistical EQG there are only the babies.)

The Hartle-Hawking
state is thus a measure $D\phi$ on $\GammaY$.
It is normalized to have total weight 1
by omitting all vacuum amplitudes from the sum ---
omitting the amplitudes of the closed $d$-manifolds.
The ensemble is the set $\cM$ of diffeomorphism classes
of $d$-manifolds with no closed connected components.
A positivity condition on the amplitudes
implies that the measure $D\phi$ is nonnegative.
So $D\phi$
is a probability measure,
the probability measure of the microscopic system.

The observables of an observer $x$ are the linear functions on 
the observer's states $\bcS\brackx$.
A linear function on $\bcS\brackx$
returns a number --- the measurement --- in each locally invariant state 
$\bphi\brackx$.
The vector space of $x$'s observables is
the dual space $\bcSx^{*}$.
The correlation functions of an experiment
are
the joint expectation values of measurements
made on the Hartle-Hawking state by the individual observers
in an experiment.
The dual of the transformation $T\colon \GammaY \rightarrow \GammaX$ maps
each macroscopic observable --- each linear function
$\bphi^{i}$ on $\GammaX$  --- to a polynomial function
$T^{*}\bphi^{i}$
on $\GammaY$.
The macroscopic correlation functions are given
by microscopic expectation values.
\eqg
\label{eq:corrfns}
T^{*}\bphi^{i}= \phi^{i}\circ T = 
\bphi^{i}_{\alpha_{1}\alpha_{2}\ldots\alpha_{n_{i}}}
\phi^{\alpha_{1}}\phi^{\alpha_{2}}\cdots\phi^{\alpha_{n_{i}}}
\qquad
\bphi^{i}\in \GammaX^{*}
\quad
\phi^{\alpha} \in \GammaY^{*}
\\[2ex]
\expval{\bphi^{i_{1}} \bphi^{i_{2}}\cdots } = \int_{\GammaX} 
\bphi^{i_{1}} \bphi^{i_{2}}\cdots  \; T_{*}D\phi
= 
\int_{\GammaY}
(T^{*}\bphi^{i_{1}}) (T^{*}\bphi^{i_{2}}) \cdots \;
D\phi
\eng
Each macroscopic observable $\bphi^{i}$ is a composite of microscopic 
observables $\phi^{\alpha}$.
Each of the two probability measures has its own partial 
entropies and total entropy.
The correlation functions --- what the observers see ---
are the observational data of the statistical theory.

The boundary of an observer is the disjoint union of connected closed
\dmo-manifolds.
The diffeomorphism classes $\Ny$ of connected closed
\dmo-manifolds are the \emph{sensors} by which
the observer makes measurements.
The sensors are in one-to-one correspondence with the microscopic 
objects.
The sensor states are exactly the microscopic states $\GammaY$.
The sensors of all the observers in an experiment
are plunged into the microscopic probability distribution.
The microscopic probability measure $D\phi$ becomes a probability 
measure on the sensor states.
The correlation functions seen by the observers
are derived from the joint expectation values of their sensors,
equation (\ref{eq:corrfns}).

$\cM$ is the set of diffeomorphism classes of compact $d$-manifolds with 
boundary that have no closed components.
$\cM$ plays two roles in statistical EQG.
$\cM$ is the ensemble that is summed over
to make the Hartle-Hawking 
state.
And $\cM$ is the set of possible experiments that can be coupled to 
the ensemble.
These are two distinct roles.
But there is a motivation, aside from mathematical suitability,
for using the set $\cM$ for both experiments and ensemble.

Consider an experiment $O\in\cM$ with the Hartle-Hawking state on its 
boundary.
Expand the Hartle-Hawking state into the sum over the ensemble.
Each diffeomorphism class $M$ in the ensemble with the same boundary as $O$
is attached to $O$ at their common boundary
to form a diffeomorphism class of \emph{closed} $d$-manifolds.
The sum over the ensemble becomes a sum over the diffeomorphism 
classes of \emph{closed} $d$-manifolds in which $O$ is embedded.
This is the ``bulk'' version of the global EQG.
The individual measurement in each embedding is locally invariant.
Global invariance is hidden in the sum over the ensemble of embeddings.
The ``bulk'' formulation of global EQG
is motivation for using $\cM$ as the set of experiments.

Actually, attaching a diffeomorphism class $O$ to a 
diffeomorphism class $M$ does not produce an individual 
diffeomorphism class of closed 
$d$-manifolds, but rather a probability measure on
the diffeomorphism classes of 
closed $d$-manifolds.
The boundary of $O$ and the boundary of $M$ are the same diffeomorphism class
of \dmo-manifolds.
Attaching $O$ to $M$ by identifying their boundaries requires 
choosing  \dmo-manifolds representing the boundary diffeomorphism class
and then choosing a diffeomorphism between the two \dmo-manifolds
to make the identification,
producing a closed diffeomorphism-class.
The globally invariant average of these gluings of $O$ to $M$
is a probability measure on the diffeomorphism classes of closed 
$d$-manifolds.
In dimensions $d\ge 3$,
the mathematics of this averaging is still to be developed
--- averaging over the infinite {mapping class group} of a closed \dmo-manifold.

There is a basic question.  In the sum over diffeomorphism classes $M$,
what numerical weight $w(M)$ is to be given to each diffeomorphism class?
A theory of EQG should provide principles that
determine the measure on the ensemble of diffeomorphism classes --- 
the weights $w(M)$.
Statistical EQG has two natural principles that together determine the weights.
The \emph{connectedness} principle simply requires that the connected
correlation functions of the statistical system 
should be given by the sum over connected $d$-manifolds.
It reduces the problem to determining the weights of the connected 
$d$-manifolds.
The second principle, the \emph{embedding} principle, is technical.
It still needs physical interpretation.
The embedding principle determines the weight of a connected 
$d$-manifold from the weights of the embedded $d$-manifolds it 
contains, i.e.~from the weights of its parts.

Expressing the embedding principle
uses technical machinery based on the probabilistic gluing of 
diffeomorphism classes of $d$-manifolds.
This machinery
might be mathematically useful as a tool 
for analyzing quantitatively the ways that $d$-manifolds 
embed in other $d$-manifolds, which is to say the ways that 
$d$-manifolds are put pieced together from  other $d$-manifolds.

The connectedness principle plus the embedding principle
completely determine the weights $w(M)$
to the extent possible ---
up to a mathematically natural set of parameters which can be absorbed into the amplitudes
as universal coupling constants of the local EQG,
for any local EQG.

This project started as an effort to understand the
``sum  over cobordisms'' calculations of
Banerjee and Moore \cite{Banerjee:2022pmw}~and
the papers they reference.
These sums over cobordisms are candidate examples of global EQG in $d=1$ and $d=2$ dimensions
where the local EQG is a topological quantum field theory.
A main impetus for constructing statistical EQG was 
to find
a principled justification for the weights used in those 
sums over cobordisms.
Banerjee and Moore's calculations gave 
several other key hints to the global structure:
a sum over $2$-cobordisms
permuting the boundary components to
produce permutation symmetry
of the boundary state,
the pervasiveness of symmetric algebras,
and a calculation that the Hilbert space norm of the Hartle-Hawking state
in the $d=2$ examples is not finite.

Mathematical rigor is not attempted here.
Convergence, regularity, etc.~are ignored.
Rigorous statistical EQG is certainly
feasible in the $d=2$ examples
since all the structural elements there are finite.
Statistical EQG
---
if it can
be constructed in $d\ge 3$ dimensions
using topological quantum field theories
as local EQG
---
might be useful in mathematics as a machine for investigating
the diffeomorphism classes of manifolds.
The strategy here is
to formulate a formally coherent (i.e.~potentially rigorous)
global structure
in arbitrary dimension $d$
when the state spaces are finite dimensional,
as when the local EQG is a topological quantum field theory.

The motivation for formulating EQG as statistical physics
is the analogy with euclidean quantum field theory.
The argument for statistical EQG is mathematical coherence.
Interpreting the Hartle-Hawking state as a quantum mechanical wave 
function is incoherent --- if only because it is not a normalizable 
state \cite{Banerjee:2022pmw}.
Statistical EQG is a mathematically coherent \emph{metaphorical} 
physical system.
It provides a coherent interpretation of the Hartle-Hawking state
as the probability measure.
The statistical system is metaphorical in the same sense as the 
statistical field theory of euclidean quantum field theory.
The statistical field theory is in four euclidean dimensions.
There are no real four dimensional observers measuring its correlation 
functions.
The statistical field theory is metaphorical.

Missing from statistical EQG is an analytic continuation to a theory 
of real time Quantum Gravity,
where the observables of the statistical theory
analytically continue to real time observables.
As it stands, the ``observers'' in statistical EQG
are metaphorical observers, like the four dimensional observers
of euclidean quantum field theory.

Beyond determining the weights $w(M)$,
it is not clear how else to test the theory.
The ``experiments'' of statistical EQG
are thought experiments in imaginary time.
A possibility of analytic continuation 
to real time is suggested in the final section below,
but even if the suggestion should turn out to be mathematically 
coherent
there is still the problem of actually constructing a local EQG
that could be tested in real time.
The only other tests would seem to be mathematical naturalness
and coherence as a theory of statistical physics.

Appendix~\ref{app:summary} summarizes the global structure.

\newpage
\section{Manifolds}

All manifolds are real, compact and unoriented.
All \dmo-manifolds are closed, i.e.~without boundary.
All state spaces are real vector spaces.
This is the setting for \emph{real} EQG.
A slightly more complicated theory ---
\emph{complex} EQG --- lives on oriented manifolds
with complex state spaces.
The term `diff-class' is used as abbreviation for `diffeomorphism class'.
$$
\begin{aligned}
\cN & = \{\text{diff-classes of \emph{closed} \dmo-manifolds}\}\\
&\qquad\qquad N \text{ is a diff-class, } \tN \text{a specific \dmo-manifold in the diff-class}
\\[2ex]
\cM & =\left\{\text{diff-classes of $d$-manifolds 
with boundary with no 
closed connected components}\right\}
\\
&\qquad\qquad M \text{ is a diff-class, }\tM \text{ a specific $d$-manifold in the diff-class}
\\[1ex]
\cM_{\emptyset} & =\left\{\text{diff-classes of \emph{closed} 
$d$-manifolds, $\partial M_{0} = \emptyset$}\right\}
\\
&\qquad\qquad M_{0} \text{ is a diff-class, }\tM_{0} \text{ a specific $d$-manifold in the diff-class}
\\[1ex]
\cM'& = \left\{\text{diff-classes of 
\emph{all} $d$-manifolds with boundary}\right\} = \cM_{\emptyset} \times \cM
\\
&\qquad\qquad M' \text{ is a diff-class, }\tM' \text{ a specific $d$-manifold in the diff-class}
\\[2ex]
\cY&= \{\Ny\}  = \{\text{diff-classes of \emph{connected closed} \dmo-manifolds}\}
\\
&\qquad\qquad\qquad\qquad y  \text{ indexes the set } \cY
\\[1ex]
\cX &= \{M\brackx\}  = \left\{\text{diff-classes of \emph{connected} 
$d$-manifolds with nonempty boundary, }
\right\}
\\
&\qquad\qquad\qquad\qquad x  \text{ indexes the set } \cX
\end{aligned}
$$
The empty manifold $\emptyset$ is a member of all four sets 
$\cN$, $\cM$, $\cM_{\emptyset}$, and $\cM'$.
Every member of $\cM'$ is uniquely the disjoint union of a member 
of $\cM_{\emptyset}$ with a member of $\cM$
so
\eq
\cM' = \cM_{\emptyset} \times \cM
\en
All four sets 
$\cN$, $\cM$, $\cM_{\emptyset}$, and $\cM'$
are semi-groups with identity.
Addition is disjoint union of manifolds.
The identity element is the empty manifold $\emptyset$.
A semigroup with identity is called a \emph{monoid}.
The boundary operator respects disjoint union (i.e.~is a morphism of 
monoids).
\eqg
\partial\colon \cM' \rightarrow \cN
\qquad
\partial \cM_{\emptyset} = \{\emptyset\}
\\[2ex]
\partial(M'_{1}\sqcup M'_{2}) = \partial M'_{1} \sqcup \partial M'_{2}
\qquad
\partial \emptyset = \emptyset
\eng
The general diff-class in $\cM$ is parametrized by the
multiplicities of connected components
\eqg
\cM = \{M^{\bm}\} 
\qquad
M^{\bm} =\sqcup_{x} \sqcup^{\bm\brackx} M\brackx
\qquad
M^{0} = \emptyset
\\[2ex]
\bm\brackx \in \NN = \{0,1,2,\ldots\}
\qquad
\textsum_{x} \bm\brackx<\infty
\eng
which is to say that $\cM$  is the free monoid on the generator 
set $X$.
Likewise, $\cN$ is the free monoid on the generator set $Y$.
\eqg
\cN = \{N^{\bn}\}
\qquad
N^{\bn} = \mathop\sqcup\limits_{y\in \cY} \sqcup^{\bny} \Ny
\qquad
N^{\mathbf{0}} = \emptyset
\\[2ex]
\bn\bracky \in \NN = \{0,1,2,\ldots\}
\qquad
\textsum_{y} \bn\bracky<\infty
\eng

The distinction between manifold $\tilde M$ and diffeomorphism 
class $M$ is key.
The tilde on the symbol for a manifold
indicates the arbitrary choice
within the diff-class.
Physics takes place on a $d$-manifold $\tM$,
invariant under diffeomorphisms of $\tM$.
Nothing physical depends on the choice of $\tM$
within the diff-class $M$.
The collection of manifolds is a \emph{class}.  The
collection of diff-classes is a \emph{set}.  Measures live on
sets, not on classes.  Summing over $d$-manifolds is done
in the set of diff-classes.

\section{Local EQG}
\subsection{State spaces and amplitudes}

The local EQG associates
to every closed \dmo-manifold $\tN$
a real vector space of states $\tcS(\tN)$.
\eq
\tN \;\;\mapsto\;\; \tcS(\tN)
\qquad
\emptyset \mapsto \Reals
\en
And it associates to every $d$-manifold
a state on the boundary,
the \emph{amplitude}.
\eq
\tM' \;\;\mapsto\;\; \tA(\tM') \qquad
\tA(\tM')\in \tcS(\partial\tM')
\en
The states and amplitudes are assumed to satisfy three
axioms which will be introduced as they are used.
Topological quantum field theories are examples of local EQG.
The three axioms of local EQG can be used as axioms for topological quantum 
field theory.

\subsection{Local diffeomorphism invariance}
The first axiom requires the amplitudes to be 
diffeomorphism invariant.
Every diffeomorphism between \dmo-manifolds
$f\in \Diff(\tN_{1},\tN_{2})$  is associated to
a linear isomorphism $L(f)$ between the state spaces.
\eqg
\tN_{1} \xrightarrow{f} \tN_{2}
\qquad\mapsto\qquad
\tcS(\tN_{1}) \xrightarrow{L(f)} \tcS(\tN_{2})
\\[2ex]
L(f_{1}\circ f_{2}) = L(f_{1})L(f_{2})
\qquad
L(1) = 1
\eng
A diffeomorphism $F \in \Diff(\tM'_{1},\tM'_{2})$  between $d$-manifolds restricts to a diffeomorphism 
$\partial F$ between the boundaries.
\eq
F \in \Diff(\tM'_{1},\tM'_{2})
\qquad\mapsto\qquad
\partial F = F_{/\partial\tM'_{1}} \in 
\Diff(\partial\tM'_{1},\partial\tM'_{2})
\en
The amplitude is required to be invariant
under diffeomorphisms between $d$-manifolds.
\eq
\tA(\tM'_{2})  =  L(\partial F) \tA(\tM'_{1})
\qquad
\forall F \in \Diff(\tM'_{1},\tM'_{2})
\en
In particular the amplitude on a particular $d$-manifold is invariant
under all diffeomorphisms of
the $d$-manifold acting as diffeomorphisms of the 
boundary.
\eq
\tA(\tM')  =L(\partial F)\tA(\tM')
\qquad
\forall F \in \Diff(\tM')
\en
Let $\btcS(\tM')$ be the subspace of $\Diff(\tM')$ invariant states 
on $\tM'$.
\eq
\btcS(\tM') = \{ \tilde\bphi \in \tcS(\partial\tM')\colon L(\partial 
F) \tilde\bphi = \tilde\bphi \;\;\forall F\in 
\Diff(\tM')\}
\en
If $\tM'_{1}$ and $\tM'_{2}$ are diffeomorphic $d$-manifolds
then $\btcS(\tM'_{1})$ can be identified with $\btcS(\tM'_{2})$
by any diffeomorphism between $\tM'_{1}$ and $\tM'_{2}$
because,
if $F_{1},F_{2}\in\Diff(\tM'_{1},\tM'_{2})$ 
are any two diffeomorphisms between them,
then $F_{1}$ and $F_{2}$ give the same identification of
locally invariant states.
\eq
(\partial F_{1})\tilde\bphi = (\partial F_{2})\partial ( F_{2}^{-1} F_{1}) \tilde\bphi  = 
 (\partial F_{2})\tilde\bphi
\qquad
\tilde\bphi\in \btcS(\tM'_{1})
\en
Any diffeomorphism between $\tM'_{1}$ and $\tM'_{2}$
gives the same identification between $\btcS(\tM'_{1})$ and
$\btcS(\tM'_{2})$.
The diffeomorphism equivalence class $\bphi$
is well defined
as a property of the diff-class.
The equivalence classes $\bphi$ are the locally invariant states
of the diff-class.
$\bcS(M')$ is the vector space 
of locally invariant states of the diff-class $M'$.
\eq
M'\in\cM' \quad \mapsto \quad \bcS(M')
= \left\{ \text{locally invariant states on $M'$} \right\}
\en
The state space of a closed $d$-manifold $\tM_{0}$ is $\btcS(\tM_{0})=\Reals$
and $\partial\Diff(\tM_{0})=\{1\}$ so the locally invariant states of 
$M_{0}$
are also the real numbers, $\bcS(M_{0}) = \Reals$.
The nontrivial locally invariant state spaces are the $\bcS(M)$ for 
$M\in \cM$.

For each $d$-manifold $\tM'$ in the diff-class $M'$
$\bcS(M')$ is identified with the subspace 
$\btcS(\tM')$
of $\Diff(\tM')$ invariant states on $\partial\tM'$.
\eq
\iota(\tM') \colon \;\cSd(M') \;\rightarrow\; \btcS(\tM') \;\subset 
\tcS(\partial\tM')
\en
The local amplitude is a locally invariant state $\bA(M')$ on each
diff-class $M'$.
\eq
M' \mapsto \bA(M') \in \cSd(M')
\en
The amplitudes of the $d$-manifold $\tM'$ is given by $\bA(M')$.
\eq
\tA(\tM') = \iota(\tM')\bA(M') 
\en
We assume that all
the vector spaces $\bcS(M')$ are finite dimensional,
as when the local EQG is a topological quantum field theory.

\section{Differential topology}

\subsection{Mapping class groups}
Suppose $\tM'$ is a $d$-manifold with
$\partial\tM'=\tN$.
Restricting diffeomorphisms of $\tM'$ to the boundary $\tN$ is a group homomorphism
\eqg
\partial\colon \Diff(\tM') \rightarrow \Diff(\tN)
\qquad
\partial F = F_{/\tN}
\\[2ex]
\partial(F_{1}\circ F_{2}) = (\partial F_{1})\circ (\partial F_{2})
\qquad
\partial 1 = 1
\eng
Let $\Diff_{0}(\tN)\subset \Diff(\tN)$ and $\Diff_{0}(\tM')\subset 
\Diff(\tM')$ be the connected components 
of the identity
in the two diffeomorphism groups.
The mapping class groups are the quotients
\eq
\MCG(\tN) = \Diff(\tN)/\Diff_{0}(\tN)
\qquad
\MCG(\tM') = \Diff(\tM')/\Diff_{0}(\tM')
\en
Any smooth vector field on the boundary $\tN$
extends to all of $\tM'$, so
\eq
\partial \big(\Diff_{0}(\tM')\big) = \Diff_{0}(\tN)
\en
Therefore restriction to the boundary is a morphism
of the mapping class groups.
\eq
\partial\colon \MCG(\tM') \rightarrow \MCG(\tN)
\en
In global EQG, 
the mapping class groups of $d$-manifolds $\tM'$
appear only in the form of the subgroups
$\partial\MCG(\tM') \subset \MCG(\tN)$.
The mapping class groups of the closed
$d$-manifolds play no role in global EQG.

\subsection{Relative diffeomorphism classes}

Suppose two $d$-manifolds $\tM'_{1}$, $\tM'_{2}$ have the same boundary 
$\tN$.
A \emph{relative diffeomorphism} between them is a diffeomorphism
that is the identity on 
the boundary.
\eq
F_{0} \in \Diff(\tM'_{1},\tM'_{2}) \qquad \partial F_{0} = F_{0}{}_{/\partial 
\tM'_{1}} = 1_{/\tN}
\qquad
\partial\tM'_{1}=\partial\tM'_{2} = \tN
\en
The equivalence class 
under relative diffeomorphisms
is the \emph{relative diffeomorphism class} $\hat M'$.
The boundary is a property of the relative diff-class, $\partial \hat M' = \tN$.
The set of relative diff-classes with 
boundary $\tN$ is written
\eq
\hcM'(\tN) = \big\{ \hat M' \colon \partial \hat M' = \tN\big\}
\en
Suppose $\tM'_{1}$ and $\tM'_{2}$ are $d$-manifolds in the same relative diff-class.
Let $F_{12}$ be any relative diffeomorphism between them.
Then  $F_{12}$ can be used to identify $\Diff(\tM'_{1})$ with 
$\Diff(\tM'_{2})$
\eqg
F_{1}\in \Diff(\tM'_{1})
\;\;\leftrightarrow\;\;
F_{2} = F_{12}\circ F_{1}\circ F_{12}^{-1} \in \Diff(\tM'_{2})
\qquad
\partial F_{1} =  \partial F_{2}
\eng
so the subgroup $\partial (\Diff(\tM'))\subset \Diff(\tN)$ depends only on 
the relative diff-class $\hat M'$.
Write this subgroup
\eqg
\partial \Diff(\hat M') \subset \Diff(\tN)
\eng
Likewise, the subgroup  $\partial \big(\MCG(\tM')\big)\subset \MCG(\tN)$ depends only on the 
relative diff-class.
\eq
\partial \MCG(\hat M') \subset \MCG(\tN)
\en

\subsection{Boundary twists}

For any $\tM'$ with boundary $\partial\tM' = \tN$
and any $f \in \Diff(\tN)$,
construct the \emph{twisted} $d$-manifold $f \tM'$ 
by identifying $\partial \tM'$ with $\tN$ using $f$.
The boundary stays the same.
Twisting is a group action.
\eq
\partial(f\tM')=\partial \tM' =\tN
\qquad
(f_{1}\circ f_{2} ) \tM' = f_{1}(f_{2}\tM')
\qquad
1 \tM' = \tM'
\en
$f\tM'$ is diffeomorphic to $\tM'$ because
the identity map from the interior of $\tM'$ to the interior of $f\tM'$
defines a diffeomorphism $F(f)$
which acts as $f$ on the boundary.
\eq
F(f)\colon \tM' \rightarrow f\tM'
\qquad
\partial F(f) = F(f)_{/\tN} = f
\en
If $F_{12}\colon \tM'_{1}\rightarrow \tM'_{2}$ is a relative diffeomorphism
then
\eq
F(f)\circ F_{12}\circ F(f)^{-1}\colon
f\tM'_{1}\rightarrow \tM'_{1}\rightarrow \tM'_{2} \rightarrow f\tM'_{2}
\en
is a relative diffeomorphism between $f\tM_{1}$ and $f\tM_{2}$.
So twisting gives an action of $\Diff(\tN)$  on the relative diff-classes.
\eq
\hat M' \mapsto f \hat M'
\qquad
f\in \Diff(\tN)
\en
If $f\in \partial \big(\Diff(\tM')\big)$,
i.e.~$f= F_{/\tN}$ for some $F\in \Diff(\tM')$,
then $ F(f) \circ F^{-1}$ is a relative diffeomorphism between $\tM'$ 
and $f\tM'$.  So the subgroup $\partial\big(\Diff(\tM')\big)$ acts trivially on the relative
diff-classes.  In particular $\Diff_{0}(\tN)$ acts trivially.
So  $\MCG(\tN)$ acts on the relative diff-classes,
the subgroup $\partial \MCG(\hat M')$ acting trivially.
\eqg
\hat M' \mapsto g \hat M'
\qquad
g \in \MCG(\tN)
\qquad
\hat M' \in \hcM'(\tN)
\\[2ex]
g \hat M' = \hat M'
\qquad
g\in \partial \MCG(\hat M')
\eng
If $\hat M'_{1}$ and $\hat M'_{2}$ are two relative diff-classes belonging to the 
same diff-class $M'_{1}=M'_{2}$ then for $\tM'_{1}$ in $\hat M'_{1}$ 
and $\tM'_{2}$ in $\hat M'_{2}$ there is a diffeomorphism $F_{12}:\tM'_{1} 
\rightarrow \tM'_{2}$.
Twisting by $\partial F_{12} = F_{12}{}_{/\tN}$ takes $\hat M'_{1}$ 
to $\hat M'_{2}$.
So the orbit of $\MCG(\tN)$ in the set of relative diff-classes $\hcM'(\tN)$
is the diff-class.
The quotient -- the set of orbits --- is
the set $\cM'_{N}$ of diff-classes with boundary $N$.
\eq
\cM'_{N}= \hcM'(\tN)/\MCG (\tN)
\en
The little group at $\hat M'$ is the subgroup $\partial\MCG(\hat 
M')\subset \MCG(\tN)$.

The set of all relative diff-classes is the product
\eq
\hcM'(\tN) = \cM_{\emptyset} \times \hcM(\tN)
\en
$\cM_{\emptyset}$ is the set of closed diff-classes.
$\hcM(\tN)$ is the set of relative diff-classes with no closed 
connected components.
$\MCG(\tN)$ acts trivially on $\cM_{\emptyset}$ so
it acts on $\hcM(\tN)$.
\eq
\cM_{N}= \hcM(\tN)/\MCG (\tN)
\en
$\cM_{N}$ is the set of diff-classes
with boundary $N$ and no closed components.

\section{Locally and globally invariant states}

Fix a \dmo-manifold $\tN$ and consider the $d$-manifolds $\tM'$ with 
boundary $\partial\tM'=\tN$.
The amplitude $\tA(\tM')$ is diffeomorphism invariant so it is
constant on each relative diff-class,
\eq
\tA(\tM'_{2}) = L\left(\partial F_{12}\right) \tA(\tM'_{1}) = 
\tA(\tM'_{1})
\qquad
F_{12}\in\Diff(\tM'_{1}, \tM'_{2})
\quad
\partial F_{12} =F_{12}{}_{/\tN} = 1
\en
so the amplitude is a function $\tA(\hat M')$ of the relative 
diff-class $\hat M'$.
\eq
\tA(\tM') = \tA(\hat M')
\en
Define $\tcS_{0}(\tN)$ to be the subspace of $\Diff_{0}(\tN)$ invariant 
states in $\tcS(\tN)$.  
\eq
\tcS_{0}(\tN) = \big\{ \tilde\phi \in \tcS(\tN) \colon f_{0} \tilde\phi=\tilde\phi \;\;\forall 
f_{0}\in \Diff_{0}(\tN)\big\}
\en
$\tcS_{0}(\tN)$ is a representation of the mapping class group
\eqg
G = \MCG(\tN)=\Diff(\tN)/\Diff_{0}(\tN)
\\[2ex]
\tilde\phi_{0} \mapsto L(g) \tilde\phi_{0}
\qquad
g\in G
\qquad
\tilde\phi_{0} \in \tcS_{0}(\tN)
\eng
The amplitude $\tA(\hat M')$ is invariant under $\Diff_{0}(\tN)$
because it is invariant under 
$\partial \Diff(\tM')$ which contains $\Diff_{0}(\tN)$ so
\eq
\tA(\hat M') \in \tcS_{0}(\tN)
\en
Let $\cS(N)$ be the subspace of $\Diff(\tN)$ invariant states in $\tcS(\tN)$
which are the same as the $\MCG(\tN)$ invariant states in $\tcS_{0}(\tN)$.
\eq
\cS(N) =\big\{ \tilde\phi_{0}\in \tcS_{0}(\tN)\colon g \tilde\phi_{0} =\tilde\phi_{0}\quad \forall g \in 
G
\big\}
\en
The subspace $\cS(N)$ is
a property of the diff-class $N$ by the usual argument.

The local subgroup
\eq
H_{\hat M'} = \partial \big(\Diff(\tM')\big) = \partial\Diff(\hat M')
\qquad
H_{\hat M'} \subset G
\en
is the same for all $\tM'$ in the relative diff-class
so the space of locally invariant states $\btcS(\hat M')$
is a property of the relative diff-class $\hat M'$.
\eq
\btcS(\hat M') = \big\{ \tilde\phi_{0} \in \tcS_{0}(\tN)\colon g \tilde\phi_{0} =\tilde\phi_{0} \quad \forall g \in 
H_{\hat M'}
\big\}
\en
The linear map $\iota(\tM')$ depends only on the relative diff-class $\hat M'$.
\eq
\iota(\hat M') \colon \cSd(M') \rightarrow \btcS(\hat M') \subset 
\tcS_{0}(\tN)
\en
and is covariant under $\MCG(\tN)$.
\eq
\iota(g\hat M') = L(g) \iota (\hat M')
\en
The globally invariant states form a subspace of the locally 
invariant states.
\eqg
\cS(N) \subset \btcS(\hat M') \subset \tcS_{0}(\tN)
\qquad
\cS(N) \subset \bcS(M')
\eng

\section{Globally invariant Hartle-Hawking state}

Recall that $\hcM(\tN)$ is the set of relative diff-classes
with no closed connected components.
The Hartle-Hawking state on $\tN$ is a sum 
of the amplitudes over $\hcM(\tN)$.
\eq
\label{eq:HH1}
D\phi(\tN) = \int_{\hcM(\tN)}  \tA(\hat M) \; \hw(\hat M) d\hat M
\qquad
D\phi(\tN) \in \tcS_{0}(\tN)
\en
$d \hat M$ is the counting measure
and $\hw(\hat M)$ is a weighting to be determined.
$D\phi(\tN)$ is in $\tcS_{0}(\tN)$  because all of the summands are.
The sum is over \emph{relative} diff-classes
because the boundary $\tN$ is fixed.
The closed $d$-manifolds are left out of the ensemble because they 
are unobservable.
The empty manifold $\emptyset$ is the only member of the ensemble
without boundary.
\eq
\hcM(\emptyset) = \{\emptyset\}
\en
The amplitude of the empty manifold is $1$.
\eq
\tA(\emptyset) = 1 \in \tcS_{0}(\emptyset) = \Reals
\en
So the normalization
\eq
D\phi(\emptyset) = 1
\en
requires assuming
\eq
\hw(\emptyset) = 1
\en
The problem is to find principles to
determine the rest of the weights $\hw(\hat M)$.

The first principle is that the Hartle-Hawking state
must be globally invariant.
$D\phi(\tN)\in \tcS_{0}(\tN)$ must be invariant under
all of $\Diff(\tN)$,
which is to say it must be invariant under all of $\MCG(\tN)$.
Global invariance ensures that the Hartle-Hawking state provides \emph{every} experiment
with a locally invariant boundary condition ---
a boundary condition
invariant under diffeomorphisms of the experiment.

Writing the amplitude in terms of the locally invariant state 
$\bA(M)$,
\eq
\tA(\hat M) = \iota(\hat M)\bA(M)
\qquad
\bA(M) \in \bcS(M)
\en
the Hartle-Hawking sum (\ref{eq:HH1}) becomes
\eq
D\phi(\tN) = \int_{\hcM(\tN)}  \iota(\hat M) \bA(M) \; \hw(\hat M) d\hat M
\en
Global invariance is 
invariance under 
$\MCG(\tN)$.
\eq
D\phi(\tN) =  L(g) D\phi(\tN) 
\qquad
\forall g\in  G=\MCG(\tN)
\en
which is
\eqa
\int_{\hcM(\tN)}  \iota(\hat M) \bA(M) \; \hw(\hat M) d\hat M&=  \int_{\hcM(\tN)} L(g) \iota(\hat M) A(M) \; \hw(\hat M) d\hat M
\\[2ex]
&=  \int_{\hcM(\tN)}\iota(g \hat M) A(M) \; \hw(\hat M) d\hat M
\\[2ex]
&=  \int_{\hcM(\tN)}\iota( \hat M) A(M) \; \hw(g^{-1}\hat M) d\hat M
\ena
The weights must be invariant under $G=\MCG(\tN)$
\eq
\hw (g\hat M ) = \hw(\hat M)
\qquad
\forall g\in G
\en
so that the Hartle-Hawking state will be globally invariant in 
general, i.e.~for every local EQG.

Write the sum over relative diff-classes 
as a sum over each orbit of $G$
followed by a sum over the set of orbits.
The set of orbits is
the set $\cM_{N}$ of diff-classes $M$ with boundary $N$ and no 
closed components.
\eq
D\phi(\tN) = \int_{\cM_{N}}  \int_{\text{ orbit } M}
\iota(\hat M)\bA(M) \; \hw( \hat M) d\hat M 
\en
The orbit of $G$ containing $\hat M$ is the homogeneous space $G/H_{\hat M}$
where  $H_{\hat M}=\{g \colon g\hat M = \hat M\}$
is the little group at $\hat M$.
The measure $\hw(\hat M) d\hat M$ 
on each orbit is $G$-invariant so
\eq
\int\limits_{\text{ orbit } M}  \iota(\hat M) \hw(\hat M) d\hat M
= P(M)w(M) dM
\en
where $w(M) dM$ is a measure on $\cM_{N}$
and $P(M)$ projects on the globally invariant states.
\eq
P(M) \colon \cSd(M) \rightarrow \cS(N)
\quad
\text{projects on } \cS(N) \subset \cSd(M)
\en
The projection is
\eq
\label{eq:iotaM}
P(M) = \int_{G} \iota(g \hat M) \;\mu d g
= \int_{G} L(g)\iota(\hat M) \;  \mu d g  
\qquad
\int_{G} \mu d g = 1
\en
where $\mu dg$ is an averaging measure, an invariant probability 
measure on $G$.
The integral (\ref{eq:iotaM}) for $P(M)$ is actually
an integral over $G/H_{\hat M}$ using the invariant probability measure $\mu dg$ pushed down to 
an invariant probability measure on $G/H_{\hat M}$.
The linear operator $P(M)$ is a property only of the manifolds and
the state spaces.  It does not depend on the amplitudes.

In dimensions $d\ge 3$,
the mapping class groups of \dmo-manifolds are 
infinite discrete groups.
The orbits $G/H_{\hat M}$ are infinite discrete homogeneous spaces.
The obvious invariant measure on $G/H_{\hat M}$ would be
the counting measure.
But this runs into a problem when
the local EQG has finite dimensional state spaces,
as when the local EQG is a topological quantum field theory.
Then the vector space $\bcS(M)$ is finite dimensional and
$\cS(N)$ is a finite dimensional subspace.
An element of $\cS(N)$ is constant on the orbit so
the sum over an infinite orbit would diverge.
The invariant measure on the orbit cannot be the counting measure or 
a multiple of the counting measure.
The invariant measure must be an averaging measure so that the integral of $\iota(\hat 
M)$ over an orbit is a projection on the subspace $\cS(N)$.

The infinite  mapping class groups $G$ are not amenable,
which is to say that there is no averaging measure on $G$ that works on 
all the bounded functions continuous with respect to the discrete topology on G.
An averaging measure is needed that works on a smaller domain of functions,
continuous with respect to a topology
for which each function $g \mapsto \tA(g\hat M)$ is continuous in $g$.
The domain of the averaging measure should be
the algebra of functions
generated by the matrix elements of the finite dimensional
(perhaps unitary) representations of $G$.
These are the bounded continuous functions in a weaker topology on 
$G$ than the discrete topology.
The averaging measure is the linear function
that projects on the matrix elements of the invariant 
states.

The globally invariant amplitude is the projection
\eq
A(M) = P(M) \bA(M)
\qquad
A(M) \in \cS(N)
\en
The Hartle-Hawking state on $\tN$ is now a sum of the globally invariant 
amplitudes over the diff-classes with boundary $N$.
Being globally invariant, it is a property of the diff-class $N$.
\eq
D\phi(N) = \int_{\cM_{N}} A(M) \; w( M) d M
\qquad
D\phi(N) \in \cS(N)
\en
The complete  Hartle-Hawking state is the sum over all diff-classes 
$N$.
\eqg
\label{eq:HHstate}
D\phi = \sum_{N\in \cN} D\phi(N) = \int_{\cM}
A(M) \; w( M) d M
\\[2ex]
D\phi \,\in\, \cS= \mathop\oplus\limits_{N\in\cN} \cS(N)
\eng
$\cS$ is the vector space of all globally invariant states.

\section{Globally invariant states as measures}

The second axiom of local EQG 
requires the state space of a disjoint union
to be the tensor product
\eq
\tN = \tN_{1}\sqcup \tN_{2}
\qquad
\tcS(\tN)
= \tcS(\tN_{1}) \otimes \tcS(\tN_{2})
\qquad
\tcS(\emptyset) = \Reals
\en
If $\{\tNc\}$ is the set of connected components of $\tN$ then
\eq
\tN = \sqcup \{\tNc\}
\qquad
\tcS(\tN)
= \mathop\otimes_{\{\tNc\}} \tcS(\tNc)
\en
The amplitude is required to be multiplicative.
\eq
\label{eq:Amultiplicative}
\tA(\tM'_{1}\sqcup \tM'_{2}) = \tA(\partial\tM'_{1}) \otimes 
\tA(\partial\tM'_{2})
\en

A diffeomorphism $f\in \Diff(\tN,\tN') $ consists of an isomorphism  between the sets of connected components
\eq
\{\tNc\}\xrightarrow{\sigma} \{\tNc'\}
\en
and a diffeomorphism between each pair of connected components.
\eq
\tcS(\tNc)
\xrightarrow{f_{\tNc}}
\tcS(\sigma \tNc)
\en
The linear action of diffeomorphisms is required
to be consistent with the tensor product.
\eq
L(f) = \mathop\otimes_{\{\tNc\}} L(f_{\tNc})
\en

Any permutation of diffeomorphic components of $\tN$
can be realized by a diffeomorphism of $\tN$.
So the space of $\Diff(\tN)$ invariant states
is the symmetric tensor product.
For the general diff-class of closed \dmo-manifolds
\eq
N^{\bn} =  \mathop\sqcup\limits_{y\in\cY} \sqcup^{\bn[y]} N[y]
\en
the space of globally invariant states is
\eq
\cS(N^{\bn}) = \mathop\otimes\limits _{y\in\cY} \otimes_{s}^{\bn[y]} \cSy
\qquad
\cSy = \cS(\Ny)
\en
The space of all globally invariant states is the symmetric algebra
on $\GammaY$.
\eqg
\label{eq:cSc}
\cS = \mathop\oplus_{N\in\cN} \cS(N) 
=  \mathop\oplus_{\bn} \cS(N^{\bn}) 
= \mathop\oplus_{\bn} \mathop\otimes\limits _{y\in\cY} \otimes_{s}^{\bn[y]} \cSy
=\mathop\oplus_{n=0}^{\infty} \left(\otimes_{s}^{n} 
\big(\oplus_{y} \cSy \big ) \right)
=\Sym\big(\oplus_{y} \cSy \big ) 
\\[2ex]
\cS=\Sym(\GammaY)
\qquad
\GammaY = \mathop\oplus_{y\in \cY} \cSy
\eng
$\GammaY$ is the space of invariant states on the connected closed 
\dmo-manifolds $\Ny$.

Choose a basis for each of the $\cSy $ and combine these bases to form a 
basis $\{\phi_{\alpha}\}$ of $\GammaY$.
Let $\{\phi^{\beta}\}$ be the dual basis of $\GammaY^{*}$,
the linear functions on $\GammaY$.
$\phi^{\beta}$ is the linear 
function on $\GammaY$ with $\phi^{\beta}(\phi_{\alpha}) = \delta_{\alpha}^{\beta}$.
The $\{\phi^{\beta}\}$ are linear coordinate functions on $\GammaY$.
The monomials $\phi_{\alpha_{1}}\phi_{\alpha_{2}}\cdots 
\phi_{\alpha_{n}}$ form a basis for the symmetric algebra 
$\cS=\Sym(\GammaY)$.
Identify $\cS$
with the convolution algebra of (generalized) measures on $\GammaY$.
\eqg
\label{eq:phis}
1 = \delta_{0}\qquad\quad
\phi_{\alpha} =(-\partial_{\alpha})\delta_{0}
\qquad\quad
\partial_{\alpha} = \frac{\partial}{\partial\phi^{\alpha}}
\\[2ex]
\phi_{\alpha_{1}}\phi_{\alpha_{2}}\cdots \phi_{\alpha_{n}}
=
(-\partial_{\alpha_{1}})(-\partial_{\alpha_{2}}) \cdots 
(-\partial_{\alpha_{n}}) \delta_{0}
\eng
$\delta_{0}$ is the Dirac delta-function measure at $0\in\GammaY$.
Multiplication is the convolution of (generalized) measures.

We conflate generalized measures with measures, in the sense that 
measures are made by summing generalized measures
as in
\eq
\delta_{\phi}= e^{\phi} = \sum_{n=0}^{\infty} \frac{1}{n!} \phi^{n}
\qquad
\phi\in\GammaY
\en
and as in summing the invariant amplitudes to make the Hartle-Hawking 
state.

\section{Hartle-Hawking state as measure}

For each diff-class $N\in\cN$ the Hartle-Hawking state
$D\phi(N)\in \cS(N)$ is a homogeneous polynomial in the basis states  
$\phi_{\alpha}$ so
a generalized measure on $\GammaY$.
The complete Hartle-Hawking state
\eq
\label{eq:Taylorseries}
D\phi = \sum_{N} D\phi(N) = \sum_{n=0} \frac1{n!} D\phi^{\alpha_{1}\ldots\alpha_{n}} 
\phi_{\alpha_{1}} \cdots \phi_{\alpha_{n}}
\en
is a formal power series in the $\phi_{\alpha}$.
The formal power series captures all the expectation values
of polynomial functions on $\GammaY$.
\eq
\label{eq:expvalues}
\langle \phi^{\alpha_{1}}\cdots \phi^{\alpha_{n}} \rangle
= \int_{\GammaY} \phi^{\alpha_{1}}\cdots \phi^{\alpha_{n}}\;D\phi
= D\phi^{\alpha_{1}\ldots\alpha_{n}} 
\en
$D\phi$ is normalized,
\eq
\int_{\GammaY} D\phi  = D\phi(\emptyset) = 1
\en
because the invariant amplitudes $A(M)$ for $M\ne\emptyset$
are derivatives of $\delta_{0}$ as measures on $\GammaY$.
We assume the positivity condition
\eq
\label{eq:globalpositivity}
\int_{\GammaY} p(\phi) ^{2}\;D\phi \;\ge\; 0
\en
for all polynomial functions $p(\phi)$.
The positivity condition on $D\phi$
should follow from
a positivity condition in the local EQG as
discussed in \cite{Banerjee:2022pmw}.

We assume that $D\phi$ is an actual measure on $\GammaY$,
that the sum of the formal measures $D\phi(N)$ over $N\in\cN$
converges to an actual measure.
$D\phi$ is then a probability measure on $\GammaY$
by the formal positivity condition and the normalization.

\subsection{The microscopic statistical system}

Interpret
\eq
\GammaY =\mathop\oplus\limits_{y\in\cY} \cSy 
\en
as the direct product of vector spaces.
An element $\phi\in\GammaY$ consists of an element $\phi\bracky\in\cSy$
for each $y\in\cY$,
i.e.~a section of the bundle $\cup_{y}\cSy \rightarrow \cY$.
The direct product is usually written $\prod_{y} \cSy$
but we are writing it additively.
The direct sum of vector spaces is a subspace of the 
direct product ---
the subspace consisting of the $\phi$ 
which vanish outside a finite subset of $\cY$.

A polynomial function on $\GammaY$ is a function
that restricts to a polynomial on the direct sum.
The formal measure $D\phi$ on $\GammaY$
is a linear function
on the polynomial functions on $\GammaY$.
We are assuming that $D\phi$ is an actual measure
--- that the sum of the formal measures $D\phi(N)$ over $N$
converges to an actual measure on $\GammaY$,
which is a probability measure
by the positivity condition and the normalization.
The ``microscopic'' statistical system consists of
\begin{itemize}[itemsep=1ex,topsep=1.5ex,leftmargin=5em]
\item the set of microscopic objects $Y=\{\Ny\}$
\item the vector space of microscopic states
$\GammaY = \mathop\oplus\limits_{y\in\cY}\, \cSy $\\[-3ex]
\item the probability measure $D\phi$ on $\GammaY$.
\end{itemize}
The vector space $\cSy$ of globally invariant states on $\Ny$
is the vector space of states of object $y$.
A microscopic state $\phi$ consists of a globally invariant 
state $\phi\bracky$ of each object $y\in\cY$.

The Fourier transform of the measure $D\phi$,
\eqa
Z(J) = \int e^{iJ} D\phi
\qquad
J= J_{\alpha}\phi^{\alpha}\in \GammaY^{*}
\ena
has Taylor series
\eq
\label{eq:ZofJ}
Z(J) = \int e^{iJ_{\alpha}\phi^{\alpha}} D\phi
= \sum_{n=0} \frac{i^{n}}{n!} J_{\alpha_{1}}\cdots J_{\alpha_{n}}
\langle \phi^{\alpha_{1}}\cdots \phi^{\alpha_{n}} \rangle
\en
$Z(J)$ is the generating function for the expectation values.
The formal measure $D\phi$ gives $Z(J)$ as a formal power series.

In $d=2$ dimensions, it should be possible to show rigorously that
$D\phi$ is a probability measure.  The circle $S^{1}$ is the only
connected closed \dmo-manifold so the microscopic system consists of a
single object $\cY = \{S^{1}\}$.
The space of states is $\GammaY= \cS{\brack{{\scriptstyle S^{1}}}}$ 
which we are assuming is a finite
dimensional vector space, as when the local EQG is a topological quantum
field theory.  If the formal power series for $Z(J)$ converges to a
continuous function rapidly enough, Bochner's theorem can be used to
show rigorously that the inverse Fourier transform $D\phi$ is an
actual probability measure on the finite dimensional vector space $\GammaY$.

\subsection{Dual cobordism symmetry}

There is an exact sequence of abelian monoids (abelian 
semigroups with identity, addition being disjoint union),
\eqg
\label{eq:exactsequenceofsemigroups}
0\rightarrow \cM_{0}
\rightarrow \cM'
\xrightarrow{\partial}
\cN
\xrightarrow{\pi}
\Omega_{d-1}
\rightarrow
0
\eng
$\Omega_{d-1} = \cN/\partial\cM'$
is the unoriented cobordism group,
an abelian group whose every nonzero element is of order 2
because $N\sqcup N = \partial(I\times N) \in \partial\cM' $.
The elements of the dual group $\hat\Omega_{d-1}$
are the characters $\chi$ of $\Omega_{d-1}$.
\eqg
\chi\colon \Omega_{d-1} \rightarrow \{\pm 1\}
\qquad
\chi(1)=1
\qquad
\chi(\omega+\omega') = \chi(\omega)\chi(\omega')
\quad
\omega,\,\omega' \in \Omega_{d-1}
\eng
For each $y\in Y$ define a representation
of the dual cobordism group on $\cSy$, 
\eqg
\chi \mapsto U\bracky (\chi)
\qquad
U\bracky(\chi) \colon \cSy \rightarrow \cSy
\qquad
U\bracky(\chi) = \chi(\pi(\Ny))
\eng
and combine these to define a representation on $\GammaY$.
\eq
\chi \mapsto U(\chi)
\qquad
U(\chi)\colon \GammaY \rightarrow \GammaY
\qquad
U(\chi) = \textstyle\prod\limits_{y} U_{y}(\chi)
\en
$\cS$ is the space of measures on 
$\GammaY$ so
the dual cobordism group acts on $\cS$
via its action on $\GammaY$.
The subspace $\cS_{\partial}$ of states invariant under 
the dual cobordism group is
the space of states on the 
diff-classes of the \emph{boundary} \dmo-manifolds.
\eq
\cS_{\partial} =  \mathop\oplus_{N \in \partial\cM'} \cS(N)
=\big\{
s \in \cS\colon\, U(\chi)_{*}s=s \quad \forall \chi \in \hat\Omega_{d-1}
\big\}
\en
$D\phi(N)=0$ when $N$ is
not a boundary so $D\phi\in\cS_{\partial}$.
The dual cobordism group is a symmetry
of the Hartle-Hawking state $D\phi$.
.

\section{Thought experiments}

\subsection{An observer as a connected $d$-manifold with non-empty 
boundary}

The experimental equipment is:

\begin{center}
\begin{tabular}{cc@{\quad}c@{}l}
\emph{sensor} & $\tNc$ &  & a connected closed
\dmo-manifold  \\[1ex]
\emph{detector} & $\tN=\sqcup\{\tNc\}$ &  & a finite set of sensors, \\
&  &  & a  closed
\dmo-manifold  \\[1ex]
\emph{observer}& $\tO_{\conn}$ &  &a connected 
$d$-manifold with non-empty boundary\\[1ex]
\emph{experiment} &  $\tO = \sqcup{\{\tO_{\conn}\}}$
& & a finite set of observers,\\
& & & a $d$-manifold with no closed connected components
\end{tabular}
\end{center}
The boundary of an observer is a detector, the disjoint union of a 
finite set of sensors.
The notation for $d$-manifolds depends on the context:
$\tM$ in the ensemble,
$\tO$ an experiment.

\subsection{Observables as dual to the locally invariant states}
An experiment $\tO$
is attached at its boundary $\tN = \partial \tO$ to 
the ensemble of $d$-manifolds.
The Hartle-Hawking state $D\phi(N)$ is
the boundary condition for $\tO$.
The possible boundary conditions on $\tO$ are the states $\tcS(\tN)$.
The observables in $\tO$  are the linear functions 
$\tilde\Phi \in \tcS(\tN)^{*}$
produced by local operations inside $\tO$.
The pairing between observable (dual state $\tilde\Phi$) and boundary 
condition (state $D\phi(N)$) is a number, the measurement of the 
observable.
The locally invariant boundary conditions are the subspace 
$\bcS(O)\subset \tcS(\tN)$.
An observable $\tilde\Phi$ is applied to a boundary condition in $\bcS(O)$
through the mapping
\eq
\tcS(\tN)^{*} \rightarrow \bcS(O)^{*}
\qquad
\tilde\Phi \mapsto \bPhi
\en
dual to the inclusion $\bcS(O) \rightarrow \tcS(\tN)$.
Observables $\tilde\Phi$ that map to the same $\bPhi$ are
equivalent --- they make the same measurements in any
locally invariant boundary condition.
$\cSd(O)^{*}$ is the space of equivalence classes of observables.

Observables are not diffeomorphism invariant.
Rather, observables that are diffeomorphism equivalent make the same measurements.
The measurements depend only on the equivalence class --- the 
locally invariant observable $\bPhi\in \cSd(O)^{*}$.
The space of observables of $O$ is $\cSd(O)^{*}$.

Each globally invariant state is locally invariant
for every local subgroup.
\eq
\cS(N) \subset \cSd(O)
\qquad
N= \partial O
\en
Restricting an observable in $\bcS(O)^{*}$ to act on the globally invariant states is dual
to the inclusion,
\eq
\cSd(O)^{*} \rightarrow \cS(N)^{*}
\qquad
\bPhi \mapsto \Phi
\en
taking the observable $\bPhi$ to 
a polynomial function $\Phi$ on $\GammaY$.
The space of all locally invariant states is
\eqg
\bcS = \mathop\oplus_{O \in \cM} \bcS(O)
\qquad \cS \subset \bcS
\qquad
\bcS^{*} \rightarrow  \cS^{*}
\qquad
\bPhi \mapsto \Phi
\eng
Measurements are expectation values
in the globally invariant Hartle-Hawking state.
\eqg
\label{eq:expvalPhi}
D\phi = \int_{\cM} A(M) \; w( M) d M
\qquad
D\phi \in \cS \subset \bcS
\\[2ex]
\langle\bPhi\rangle = \bPhi\, D\phi = \Phi\, D\phi
= \int_{\GammaY} \Phi\,D\phi
\eng
The first formula for the measurement $\langle\bPhi\rangle$
is from the point of view of the observers.
The globally invariant state $D\phi$
is a locally invariant boundary condition
experienced by the local observerable $\bPhi$.
In the second formula,
$\bPhi\in\bcS^{*}$ is transformed to the sensor observable $\Phi\in\cS^{*}$
which is evaluated on the globally invariant state $D\phi$.
The third formula is equivalent to the second.
$\Phi\in \cS^{*}$ is a function on the invariant sensor states $\GammaY$
and $D\phi\in \cS$ is the probability measure on $\GammaY$. 
The measurement is made by plunging the sensors at the boundary of $O$
into the probability distribution of microscopic states.

Observables and measurements are now associated to the diff-classes,
not the specific manifolds.
\emph{Sensors}, \emph{detectors}, \emph{observers}, and \emph{experiments}
refer to the diff-classes from now on.

\begin{center}
\begin{tabular}{cc@{\quad}c@{\;}l}
\emph{sensor} & $\Ny$ &  & a connected closed diff-class of
\dmo-manifolds  \\[1ex]
\emph{detector} & $N^{\bn}= \sqcup_{y}\sqcup^{\bn[y]}\Ny$ &  & a 
disjoint union of sensors, \\
& &  & a diff-class 
of closed \dmo-manifolds  \\[1ex]
\emph{observer}& $\Ox$ &  &a connected diff-classes of
$d$-manifolds with \\
&&&non-empty boundary, $\Ox = \Mx$ \\[1ex]
\emph{experiment} &  $O^{\bm} = \sqcup_{x} \sqcup^{\bm\brackx}
\Ox$ & & a disjoint union of observers, \\
& & &   a diff-class of $d$-manifolds with no closed \\
& & &  connected components
\end{tabular}
\end{center}
The observer $\Ox$ is the same diff-class as $\Mx$
just labeled differently.
$\cX=\{\Ox\}$ is the set of observers,
the set of macroscopic objects.

Experiments are parametrized by the multi-index $\bm[x]$
which tells how many of each observer are contained in the experiment.
Detectors are parametrized by the multi-index $\bn[y]$
which tells how many of each sensor are contained in the detector.
The boundary of each observer is a detector,
\eq
\partial \Ox  = \mathop\sqcup\limits_{y}\sqcup^{\bn[x,y]}\Ny
\en
as is the boundary of an experiment.
\eq
\partial O^{\bm} = \mathop\sqcup\limits_{x} \sqcup^{\bm\brackx}\partial \Ox
= \mathop\sqcup\limits_{y}\sqcup^{\bm\cdot\bn [y]}\Ny
\qquad
\bm\cdot\bn [y] = \mathop{\textstyle\sum}\limits_{x}\bm\brackx\,\bn[x,y]
\en

$\bcSx = \cSd(\Ox)$ is the space of locally invariant states of the observer $\Ox$.
The space of all locally invariant observer states is the direct 
product
\eq
\GammaX = \mathop\oplus\limits_{x\in\cX} \bcSx
\en
$\GammaX$ is the space of macroscopic states.
An element $\bphi\in\GammaX$ is
a locally invariant observer state $\bphi\brackx \in \cSd(\Ox)$ for each observer 
$x\in X$.
The space of locally invariant states of an experiment $O^{\bm}$ is the 
symmetric tensor product
\eq
\cSd(O^{\bm}) = \mathop\otimes\limits_{x} \otimes_{s}^{\bm\brackx} \bcSx
\en
The space of all locally invariant states is the symmetric algebra on 
$\GammaX$.
\eq
\cSd = \mathop\oplus\limits_{\bm} \cSd(O^{\bm}) 
= \mathop\oplus\limits_{\bm} \mathop\otimes\limits_{x} 
\otimes_{s}^{\bm\brackx} 
\bcSx
= \Sym(\GammaX)
\en
$\GammaX^{*}$ is the space of individual observables ---
the observables of the individual observers,
\eq
\GammaX^{*} = \mathop\oplus\limits_{x} \bcSx^{*}
\en
Each experiment has locally invariant observables
\eq
\cSd(O^{\bm})^{*} = \mathop\otimes\limits_{x} \otimes_{s}^{\bm\brackx} 
\bcSx^{*}
\en
The space of all locally invariant observables is the symmetric 
algebra on $\GammaX^{*}$.
\eq
\cSd^{*} = \mathop\oplus\limits_{\bm} \cSd(O^{\bm}) ^{*}
= \mathop\oplus\limits_{\bm} \mathop\otimes\limits_{x} 
\otimes_{s}^{\bm\brackx} 
\bcSx^{*}
= \Sym(\GammaX^{*})
\en
which is the vector space of polynomial functions on $\GammaX$.

\subsection{Measurements as correlation functions of observers}

Write the individual observable
\eq
\bJ = \textsum_{x} \bJ\brackx
\qquad
\bJ \in \bcS^{*}
\qquad
\bJ\brackx \in \bcSx^{*}
\en
The space $\bcS(O^{\bm})^{*}$ of observables in experiment $O^{\bm}$ 
is spanned by the monomials
\eq
\bJ^{\bm} = \textprod_{x} \bJ\brackx^{\bm\brackx}
\en
The measurements in $O^{\bm}$ are spanned by the correlation functions
\eq
\expval{\bJ^{\bm} } = \expval{\textprod_{x} \bJ\brackx^{\bm\brackx}}
\en
which are the moments of a probability  measure 
$D\bphi$ on 
$\GammaX$.
\eq
\expval{\bJ^{\bm} } =  \int_{\GammaX} \bJ^{\bm} D\bphi
\en
The correlation functions are the joint expectation values of 
measurements by the individual observers in $O^{\bm}$.
The correlation functions are a complete set of measurements in $O^{\bm}$.
The set of correlation functions for all the  experiments $O^{\bm}$ is the 
observational data of the statistical EQG.
The generating function of the correlation functions is
\eqg
\bZ(\bJ) = \expval{e^{\bJ}}
=
\sum_{\bn} \frac1{\bm!} \expval{\bJ^{\bm}}
=\int_{\GammaX} e^{\bJ} D\bphi
\\[2ex]
e^{\bJ} = e^{\sum_{x}\bJ\brackx}
=
\sum_{\bm} \frac1{\bm!} \bJ^{\bm}
\qquad
\bm! = \textprod_{x} \bm\brackx!
\eng
The generating function of the connected correlation 
functions is
\eq
\bW(\bJ) = \ln \bZ(\bJ)  = \sum_{\bm} \frac1{\bm!} 
\expval{\bJ^{\bm}}_{\mathrm{conn}}
\en
The connected correlation function
expresses correlations among the entire set of observers
that do not come from correlations within subsets of the observers.

There is no distinguished observer with access to the 
boundaries of $d$-manifolds.
Any principles 
must be imposed on the 
correlation functions
of
observers coupled to the 
statistical system.

\subsection{Microscopic states map to observer states}

The sensors $\Ny$ are in one-to-one correspondence with the microscopic 
objects $\Ny$.
The space of sensor states is identical to the space of microscopic 
states.
\eq
\Gammasensor =\GammaY = \mathop\oplus_{y}\, \cSy 
\en
The sensor state \emph{is}  the microscopic state
(or the microscopic state \emph{is} the sensor state).
The space of observer states
is the direct product
of the locally invariant state spaces.
\eq
\Gamma_{\obs} = \GammaX =\mathop\oplus_{x}\bcSx
\en
The observer state \emph{is}  the macroscopic state
(or the macroscopic state \emph{is} the observer state).

Write the sensor states
\eq
\phi\in\GammaY
\qquad
\phi = \textsum_{y} \phi\bracky
\qquad
 \phi\bracky \in \cSy
\en
and the observer states
\eq
\bphi \in \GammaX
\qquad
\bphi = \textsum_{x} \bphi\brackx
\qquad
\bphi\brackx \in \bcSx
\en
Every globally invariant state is locally invariant
so there is a function, a transformation,
from sensor states (= microscopic states) to observer states.
\eq
\GammaY \;\xrightarrow{T}\; \GammaX
\qquad
\phi \xmapsto{T} \bphi = T(\phi)
\en
Each $\bphi\brackx$ is a homogeneous polynomial in the 
$\{\phi[y] \colon \Ny\subset \partial \Ox\}$.

$T_{*}$ maps
measures on $\GammaY$ to measures on $\GammaX$.
\eq
\cS \xrightarrow{T_{*}} \bcS
\qquad
T_{*} \delta_{\phi} = \delta_{T(\phi)}
\qquad
D\bphi = T_{*}D\phi
\en
The probability measure $D\phi$ on the sensor states
is mapped to the probability measure $D\bphi$ on the observer states.
The pullback $T^{*}$ maps
functions on $\GammaX$ to functions on $\GammaY$.
\eq
\cSd^{*} \xrightarrow{T^{*}} \cS^{*}
\qquad
\Phi =T^{*} \bPhi  = \bPhi\circ T
\en
The individual observable $\bJ\brackx\in\bcSx$ is 
mapped to a homogeneous polynomial on 
$\GammaY$
of total degree the number of connected components in $\partial\Ox$.
\eq
T^{*}\bJ\brackx = \frac1{n(x)!}\, T^{*}\bJ\brackx_{\alpha_{1}\ldots \alpha_{n(x)}}
\,\phi^{\alpha_{1}}\cdots \phi^{\alpha_{n(x)}}
\qquad
n(x) = \textsum_{y}  \bn[x,y]
\en
The correlation function
\eq
\expval{\bJ^{\bm} } 
= \int_{\GammaX} \bJ^{\bm} D\bphi
= \int_{\GammaY} T^{*}\bJ^{\bm} D\phi
=
\int_{\GammaY} \textprod_{x} \big(T^{*}\bJ\brackx\big)^{\bm\brackx}\,  D\phi
\en
is the expectation value of a polynomial function on $\GammaY$.
Each individual observable $\bJ\brackx$ on $\GammaX$ is a composite observable
$T^{*}\bJ\brackx$ on $\GammaY$.

\subsection{Entropies}

The microscopic state space is the direct product
\eq
\GammaY = \Gammasensor = \mathop\oplus_{y\in \cY} \cSy
\qquad
\cY = \big\{\Ny\big\}
\en
The macroscopic state space is the direct product
\eq
\GammaX = \Gamma_{\obs} = \mathop\oplus_{x\in \cX} \bcSx
\qquad
\cX = \big\{\Mx\big\}
\en
Each microscopic state gives a macroscopic state.
\eq
\GammaY \xrightarrow{T}\GammaX
\en
The microscopic statistical system is the state space $\GammaY$ with 
probability measure $D\phi$.
The macroscopic system is the state space $\GammaX$ with
probability measure $D\bphi = T_{*}D\phi$.
Both are composite systems.  Each component --- each object --- can be thought of as a 
property.  Microscopic property $y$ is parametrized by $\cSy$.
Macroscopic property $x$ is parametrized by $\bcSx$.
Sensor $\Ny$ registers microscopic property $y$.
Observer $\Ox$ registers macroscopic property $x$.

Each layer has its marginal entropies.
For every finite subset $V\subset \cY$ 
there is a subsystem $\Gamma_{V} = \mathop\oplus_{y\in V} \cSy$.
The projection $\GammaY \rightarrow \Gamma_{V}$
takes $D\phi$ to the marginal probability measure $D\phi|V$
on $\Gamma_{V}$.  
Every expectation value of a polynomial in the $\phi\bracky$
is a moment of the marginal measure on the subset $V$ of
sensors $y$ occuring in the polynomial.
The marginal entropy in $V$ is
\eq
H(V) = -\int_{\Gamma_{V}} 
\ln \left(
\frac{D\phi|V}{\prod_{y\in V} D\phi|\{y\}}
\right)
D\phi|V
\en
The total entropy $H(\GammaY)$ is the limit over all finite subsets $V$.
The marginal entropies $H(V)$ determine 
all the joint and conditional entropies.

In the same way, every finite subset $U \subset \cX$
defines a macroscopic subsystem $\Gamma_{U} = \mathop\oplus_{x\in \bU} \bcSx$
with marginal probability measure $D\bphi|U$.
The correlation functions
are the moments of these marginal measures.
The marginal entropy is
\eq
H(U) = -\int_{\Gamma_{U}} 
\ln \left(
\frac{D\bphi|\bU}{\prod_{x\in \bU} D\bphi|\{x\}}
\right)
D\bphi|U
\en
The limit over finite subsets $U$ is the total macroscopic observer entropy 
$H(\GammaX)$.

\section{Connectedness principle}

The weights $w(M)$  in the  sum over diffeomorphism classes
are still arbitrary.
We need to impose  principles on the statistical EQG
that will determine the weights $w(M)$.
Useful principles might be expected
to be related to properties of the local $\text{EQG}$.
The elements of local EQG that were used in
formulating statistical EQG 
were the diffeomorphisms acting on the state spaces,
the tensor product structure of the state spaces,
and the diffeomorphism invariance of the amplitudes.
The multiplicative property of the amplitude,
equation (\ref{eq:Amultiplicative}),
was not used.

$\cX\subset \cM$  is the set of 
diff-classes $\Mx$ of \emph{connected} $d$-manifolds with non-empty 
boundary.
Let $D\phic$ be the measure on $\GammaY$  obtained by summing over 
$X$.
\eq
D\phic = \int_{\cX} A(M) w(M) dM
\en
The \emph{connectedness principle} is the equation
\eq
\label{eq:connectedness}
D\phi = \exp (D\phic)
\en
where exponentiation
is in the commutative convolution algebra $\cS$
of measures on $\GammaY$.

The connected expectation values in the probability distribution 
$D\phi$ on 
$\GammaY$ are generated by the logarithm of the generating 
function/Fourier transform 
$Z(J)$
defined in equation (\ref{eq:ZofJ}).
\eq
W(J) = \ln Z(J) 
= \sum_{n=0} \frac{i^{n}}{n!} J_{\alpha_{1}}\cdots J_{\alpha_{n}}
\langle \phi^{\alpha_{1}}\cdots \phi^{\alpha_{n}} 
\rangle_{\mathrm{conn}}
\en
The function $Z(J)$ on $\GammaY^{*}$ is the Fourier transform of the 
measure $D\phi$ on $\GammaY$.
The Fourier transform takes convolution of measures to multiplication
of functions.
So the connectedness principle is equivalent to
\eq
W(J) = \int_{\cX} e^{iJ} D\phic
\qquad
\langle \phi^{\alpha_{1}}\cdots \phi^{\alpha_{n}} 
\rangle_{\mathrm{conn}}
= \int_{\cX}\phi^{\alpha_{1}}\cdots \phi^{\alpha_{n}}  D\phic 
\en
which is to say that the connected expectation values in the 
probability distribution $D\phi$ are given
by summing over the diff-classes of \emph{connected} $d$-manifolds.
The macroscopic correlation functions are expectation values of 
composite microscopic observables,
so the connectedness principle ensures that all the connected correlation 
functions of observers are the sums over connected diff-classes.

The multiplicative property (\ref{eq:Amultiplicative}) of the amplitudes
is equivalent to the condition that
the locally invariant amplitude
\eq
\bA\colon \cM \rightarrow \bcS
\en
is a representation of the monoid,
\eq
\bA(M_{1}\sqcup M_{2}) = \bA(M_{1}) \bA(M_{2})
\qquad
\bA(\emptyset) = 1
\en
The projection $P(M)\colon \bcS(M) \rightarrow \cS$ on globally invariant states is multiplicative
\eq
P(M_{1}\sqcup M_{2})(\bphi_{1} \bphi_{2}) = [P(M_{1}) \bphi_{1}] [ 
(P(M_{2}) \bphi_{2}]
\en
so the globally invariant amplitude $A(M) = P(M) \bA(M)$ is multiplicative
\eq
A\colon \cM \rightarrow \cS
\qquad
A(M_{1}\sqcup M_{2}) = A(M_{1}) A(M_{2})
\qquad
A(\emptyset) = 1
\en
The measures on $\cM$ form a commutative algebra under convolution
\eq
\delta_{M_{1}}\delta_{M_{2}} = \delta_{M_{1}\sqcup M_{2}}
\en
$A\colon \cM \rightarrow \cS$ extends to a function $A_{*}$ that 
takes  a measure on $\cM$ to $\cS$.
\eq
A_{*}\colon \meas(\cM) \rightarrow \cS
\qquad
A_{*}(\delta_{M}) = A(M)
\en
which is a morphism of commutative algebras.
\eq
A_{*}(\delta_{M_{1}}\delta_{M_{2}}) = 
A_{*}(\delta_{M_{1}\sqcup M_{2}}) = A(M_{1}\sqcup M_{2}) = A(M_{1}) A(M_{2})
= A_{*}(\delta_{M_{1}}) A_{*}(\delta_{M_{2}}) 
\en
The Hartle-Hawking state is the measure $w dM$ on $\cM$
mapped by the invariant amplitude to an invariant state.
\eq
D\phi = A_{*}(wdM)
\en

The connectedness property (\ref{eq:connectedness}) will hold for
every local EQG if
\eq
\label{eq:connectednessM}
wdM = \exp(wdM_{\conn})
\en
for $wdM_{\conn}$ a measure on $\cX$.
Then
\eqg
D\phi = A_{*}(wdM) = A_{*}(\exp(wdM_{\conn}))
=  \exp(A_{*}(wdM_{\conn}))
= \exp(D\phi_{c})
\eng
with
\eq
D\phi_{c}=A_{*}(wdM_{\conn})
\en
$\cX=\{\Mx\}$ is the set of connected diff-classes with non-empty 
boundary.
The general diff-class in $\cM$ is parametrized by a multi-index 
$\bm\brackx$.
\eq
M^{\bm} = \mathop\sqcup\limits_{x} \sqcup^{\bm\brackx} \Mx
\en
Write the measure on $\cX$
\eq
wdM_{c} = \textsum_{x} \bw\brackx \delta_{x}
\qquad
\bw\brackx= w(\Mx)
\en
Then
\eqg
\exp(wdM_{c}) = \sum_{\bm} \frac{\bw^{\bm}}{\bm!}  dM^{\bm}
\\[2ex]
\bw^{\bm} = \textprod_{x} \bw\brackx^{\bm\brackx}
\qquad
dM^{\bm} = \textprod_{x} \delta_{x}^{\bm\brackx}
\eng
while
\eq
w dM = \sum_{\bm} w(M^{\bm})  dM^{\bm}
\en
Equation (\ref{eq:connectednessM}) becomes
\eq
\label{eq:factorizationw}
w(M^{\bm}) =\frac{\bw^{\bm}}{\bm!} 
\en
The connectedness principle (\ref{eq:connectednessM}) is satisfied for every local EQG
if $w(M)$ satsifies the factorization condition (\ref{eq:factorizationw}),

The weights $\bw\brackx$ of the 
connected diff-classes $\Mx$
remain to be determined.

\section{Gluing axiom of local EQG}
\label{sect:GluingAxiom}

The third and last axiom of local EQG is the gluing axiom.
Each state space $\tcS(\tNc)$ is endowed with a positive symmetric bilinear form
\eq
K(\tNc) \in \big[ \tcS(\tNc) \otimes_{s} \tcS(\tNc)\big]^{*}
\en
$K(\tNc)$ can also be regarded as a linear function from states to 
dual states.
\eq
\tcS(\tNc) \xrightarrow{K(\tNc)} \cS(\tNc)^{*}
\qquad
\tA \mapsto \tA^{t} = \tA K(\tNc)
\qquad
\tA_{1}  K(\tNc)\tA_{2} = \tA_{2}  K(\tNc)\tA_{1}
\en
$K$ is requived to be invariant under diffeomorphisms.  
\eq
K(\tNc) = L(f)^{*} \circ K(\tNc') \circ L(f)
\qquad f\in \Diff(\tNc,\tNc')
\en
A positive symmetric bilinear form $K(\tN)$
is associated to the general \dmo-manifold $\tN$
by tensoring the bilinear forms of the individual connected 
components of $\tN$.

The bilinear forms $K(\tN)$
played no role in the construction of statistical {EQG}{}.
They would have been needed 
to construct
the invariant state space
as a Fock space of square-integrable quantum states.
But the invariant state 
space was constructed as the space of measures on a vector space.
The bilinear forms were not needed.

Suppose $\tM$ is a $d$-manifold whose boundary
contains two connected components
$\tN_{\conn,{1}}$ and
$\tN_{\conn,{2}}$ which are diffeomorphic,
and suppose $f_{12}$ is any diffeomorphism between them.
\eq
\partial \tM = \tN = \tN_{\conn,{1}}\sqcup \tN_{\conn,{2}} \sqcup \tN'
\qquad
f_{12} \in \Diff(\tN_{\conn,{1}}, \tN_{\conn,{2}})
\en
Let $\tM'$ be the $d$-manifold with boundary $\tN'$ formed by identifying 
$\tN_{\conn,{1}}$
with $\tN_{\conn,{2}}$ using $f_{12}$.
Let $K_{12}$ be the bilinear form between the two state spaces.
\eq
K_{12} = K(\tN_{\conn,{2}}) \circ L(f_{12})
\in 
\left[\tcS(\tN_{\conn,{1}}) \otimes_{s} \tcS(\tN_{\conn,{2}}) \right]^{*}
\en
Contracting with $K_{12}$ takes states on $\tN=\partial \tM$ to states 
on $\tN' = \partial \tM'$.
\eq
K_{12}  \iprod \tA(\tM) \in \tcS(\tN')
\en
Contracting with $K_{12}$ is summing over the local EQG states
passing through $\tN_{\conn,2} = f_{12}\tN_{\conn,1}$.
The gluing axiom is
\eq
K_{12}  \iprod \tA(\tM) = \tA(\tM')
\en
If $\tM$ is a $d$-manifold whose boundary contains two 
\dmo-manifolds which are diffeomorphic,
and $f_{12}$ is any diffeomorphism between them,
\eq
\partial \tM = \tN = \tN_{{1}}\sqcup \tN_{{2}} \sqcup \tN'
\qquad
f_{12} \in \Diff(\tN_{1}, \tN_{2})
\en
then repeated use of the gluing axiom on the connected components of 
$\tN_{1}$ and $\tN_{2}$ gives a gluing formula
\eq
K_{12}  \iprod \tA(\tM) = \tA(\tM')
\en
where $\tM'$ is the $d$-manifold formed by identifying 
$\tN_{1}$ and $\tN_{2}$ with $f_{12}$
and $K_{12}$ is the quadratic form on their state spaces
made with $f_{12}$.

\section{Quiescent experiments}

A sum over states occurs
in statistical EQG
only
in the coupling between an experiment $O$ and
a diff-class $M$ in the ensemble.
The gluing axiom will come into play 
when the observable on $\tO$
is the transposed amplitude $\tA(\tO)^{t}$.
Call this the \emph{quiescent} observable of the experiment $O$.
The diffeomorphism invariant bilinear form $K(\partial\tO)$
restricts to an invariant bilinear form $\bK(O)$ on the locally invariant 
states $\bcS(O)$.
\eq
\bK(O)\colon \bcS(O)\rightarrow \bcS(O)^{*}
\en
The quiescent observable is the locally invariant dual state
\eq
\bQ(O) = \bA(O)^{t}=\bA(O) \bK(O) 
\qquad
\bQ(O)\in \bS(O)^{*}
\en
$\bQ(O)$ is a natural observable associated to the diff-class $O$.

$\bK(O)$ restricts to a bilinear form on the globally invariant 
states $\cS(\partial O)$
\eq
K(O)\colon \cS(\partial O)\rightarrow \cS(\partial O)^{*}
\en
The quiescent observable $\bQ(O)$ projects to
a function on the microscopic state space $\GammaY$
\eq
Q(O) = T^{*} \bQ(O) = A(O)^{t} = A(O) K(\partial O)
\qquad
Q(O) \in \cS(\partial O)^{*}
\en
The bilinear form respects tensor products so
$O\mapsto \bQ(O)$ is multiplicative.
\eqg
\bQ(O\sqcup O') = \bQ(O) \bQ(O')
\qquad
\bQ(\emptyset) = 1
\\[2ex]
\bQ(O^{\bm}) = \bQ^{\bm} = \prod_{x} \bQ\brackx^{\bm\brackx}
\qquad
\bQ\brackx = \bQ(\Ox) \in \bcSx^{*}
\eng
Likewise, $O\mapsto Q(O)$ is multiplicative.
\eq
Q(O^{\bm}) = Q^{\bm} = \prod_{x} Q\brackx^{\bm\brackx}
\qquad
Q\brackx = Q(\Ox)
\en
$\bQ\brackx$ is the quiescent observable of observer $x$,
a linear function on $\bcSx$.
$Q\brackx$ is its globally invariant projection.
Introduce real variables $\boldsymbol{q}\brackx$ and
define
\eqg
\bq{\cdot} \bQ
= \int_{\cX}\bq \bQ \, \bw dx
= \sum_{x} \bq\brackx\bQ\brackx \bw\brackx
\qquad
\bq\brackx\in \Reals
\\[2ex]
e^{\bq{\cdot} \bQ}
=
e^{\int_{X} \bq\bQ\,\bw dx}
= \prod_{x\in\cX} e^{\bq\brackx\bQ\brackx\bw\brackx}
=\sum_{\bm} \frac{\bw^{\bm}\bq^{\bm} }{\bm!}\bQ^{\bm}
\eng
Call $e^{\bq{\cdot} \bQ}$ \emph{the} quiescent observable --- the 
generating function for all the quiescent experiments.

There is an analogy to euclidean quantum field theory/statistical field theory.
$\cX$ is the ``spacetime'' on which the fields live.
$\bcSx$ is the space of classical degrees of freedom at $x$.
The classical field is $\bphi$
which takes values $\bphi\brackx\in \bcSx$.
The space of fields is the direct product
$\GammaX = \mathop\oplus_{x} \bcS_{x}$.
The probability measure $D\bphi$ on $\GammaX$
is analogous to the normalized path-integral measure
on fields in euclidean quantum field theory.
$\bQ\brackx$ is a local observable.
$e^{\int_{X} \bq\bQ\,\bw dx}$ 
is a mathematically natural source term.

The conditional correlation functions are
\eq
\expval{\bPhi}_{|\bq} = \frac{
\expval{\bPhi\; e^{\bq{\cdot} \bQ}}
}
{\expval{ e^{\bq{\cdot} \bQ}}}
= \frac{
\displaystyle\int_{\GammaX} \bPhi\, e^{\int_{\cX}\bq \bQ \, \bw dx}\, D\bphi
}
{\displaystyle\int_{\GammaX}  e^{\int_{\cX}\bq \bQ \, \bw dx}\, D\bphi
}
\en
where $\bPhi$ is an arbitrary observable.
They are conditional microscopic expectation values
\eq
\expval{\bPhi}_{|\bq} = \frac{
\expval{\Phi\; e^{\bq{\cdot} Q}}
}
{\expval{ e^{\bq{\cdot} Q}}}
= \frac{
\displaystyle\int_{\GammaY} \Phi\, e^{\int_{\cX}\bq Q \, \bw dx}\, D\phi
}
{\displaystyle\int_{\GammaY}  e^{\int_{\cX}\bq Q \, \bw dx}\, D\phi
}
\en
which are expectation values in the conditional
probability measure on $\GammaY$
\eq
D\phi_{|\bq} = \frac{
e^{\bq{\cdot} Q}\, D\phi
}
{\expval{ e^{\bq{\cdot} Q}}}
\en

\section{Bulk representation of the conditional measure}

Now we express the conditional measure $D\phi_{|\bq}$ as a sum over 
diff-classes of $d$-manifolds in which the quiescent experiments are embedded.
Let
\eqa
\cX_{0} &=\{\text{closed connected diff-classes}\}
\\[2ex]
\cX &=\{\text{connected diff-classes with non-empty boundary}\}
\\[2ex]
\cX' &= \cX_{0}\cup\cX = \{\text{all connected diff-classes}\}
\ena
\eq
\begin{array}{c@{\;=\;}c@{}ll}
\cM_{0} & \mathop\oplus\limits_{x_{0}\in\cX_{0}} &\NN x_{0} &= \{\text{closed diff-classes}\}
\\[2.5ex]
\cM & \mathop\oplus\limits_{x\in\cX} &\NN x &= \{\text{diff-classes with no closed components}\}
\\[2.5ex]
\cM' &  \mathop\oplus\limits_{x'\in\cX'}& \NN x' &= \{\text{all diff-classes}\}
\end{array}
\en
Now $\cM$, $\cM_{0}$, and $\cM'$ are the direct products, not the direct 
sums.  An actual diff-class has a finite number of connected 
components --- it lies in the direct sum.
But measures live on the direct product.
The sum over diff-classes is the formal power series 
of an integral with respect to a measure on the direct 
product $\cM$.
The measure on $\cM$ is
\eqg
wdM =e^{\int_{\cX} \delta_{x} \bw dx}
= \sum_{\bm} \frac{\bw^{\bm}}{\bm!} \prod_{x} \delta_{x}^{\bm\sbrack{x}}
= \sum_{\bm} \frac{\bw^{\bm}}{\bm!} \delta_{M^{\bm}}
\eng
The weights $\bw\brackx$ are still arbitrary.
The locally invariant amplitudes are encoded in the element
\eq
\bA \in \GammaX
\qquad
\bA[x] = \bA(\Mx)\in \bcSx
\en
$P$ is the projection on globally invariant states
(by averaging over the mapping class groups).
\eq
P \colon \bcS \rightarrow \cS
\qquad
P_{/\cS} = 1
\en
The Hartle-Hawking state is
\eq
D\phi = P \bA_{*}\, e^{\int_{\cX} \delta_{x} \bw dx}
= P \, e^{\int_{\cX}  \bA \bw dx}
= e^{\int_{\cX}  P \bA \bw dx}
=  e^{\int_{\cX}  A  \bw dx}
\en
where $x\mapsto A\brackx$ is the globally invariant amplitude,
\eq
A = P \bA
\qquad
A\brackx = P \bA(\bM\brackx)  \in \cS
\en
Each $A\brackx$ is a polynomial state in $\cS$.

Up to this point there has been no need for weights on the closed 
diff-classes or for amplitudes on the closed diff-classes.
Now choose weights $\bw_{0}\brack{x_{0}}$ on the connected closed
diff-classes $x_{0}\in \cX_{0}$ (arbitrarily for now).
The weights on $X'=X_{0}\cup X$ are $\bw'=(\bw_{0},\bw)$.
The locally and globally invariant amplitudes $\bA$ and $A$ are defined on $X_{0}$ so on $\cX'$.
\eqg
\bA' = (\bA_{0},\bA)
\qquad
A' = (A_{0},A)
\qquad
A_{0}\brack{x_{0}} = \bA_{0}\brack{x_{0}} = \bA(M_{0}\brack{x_{0}})
\eng
The amplitude $\bA(M_{0}\brack{x_{0}})$ of a closed diff-class $M_{0}\brack{x_{0}}$
is a number so
$A_{0}\brack{x_{0}} = \bA_{0}\brack{x_{0}}$.

The gluing axiom of local EQG implies an identity
\eq
\label{eq:invgluingformula}
A(O)^{t} A_{*} = A_{*} O_{\couple}
\en
which is
\eq
\qquad
A(O)^{t} A(M') = A_{*}(O_{\couple}\delta_{M'})
\qquad
M' \in \cM'
\en
where
$O_{\couple}$ is a linear operator on measures on $\cM'$,
the \emph{embedding operator}
\eqa
O_{\couple}&\,\colon\, \meas(\cM') \rightarrow \meas(\cM')
\\[3ex]
O_{\couple} \delta_{M'}
&=
\left\{
\begin{array}{c@{\qquad}l}
\displaystyle\int_{\MCG(\partial\hat M')}
\delta_{E(\hat O, g \hat M')}
\,\mu dg
&
\partial O \subseteq \partial M'
\\[3ex]
0 & \partial O \not\subseteq \partial M'
\end{array}
\right .
\ena
If $\partial O$ is contained in $\partial M'$,
choose a representative  $\tN'$ of 
$\partial M'$ and a representative $\tN_{O}$ of $\partial O$ such that
$\tN_{M'} = \tN_{O} \sqcup \tN$
for some $\tN$.
Then choose relative diff-classes $\hat O$ and
$\hat M'$ representing $O$ and $M'$.
Then, for each $g\in \MCG(\tN')$
form the diff-class $E'(\hat O, g \hat M')$ with  boundary $N$
by identifying $\tN_{O} = \partial 
\hat O$ with $\tN_{O} \subset \partial (g \hat M')$.
Then average the delta-function at $E'\in \cM'$ 
over the mapping class group.
Averaging over the mapping class group removes all dependence on 
the arbitrary choices.

The embedding operator $O_{\couple}$ takes the diff-class $M'\in \cM'$
to a probability measure on $\cM'$ supported on the diff-classes $E'$ which
contain $O$ embedded in the interior of $E'$ with complement $M'$.
The map $O \mapsto O_{\couple}$ is multiplicative
\eq
(O_{1}\sqcup O_{2})_{\couple} = O_{1\couple}O_{2\couple} = O_{2\couple} O_{1\couple}
\en
so the algebra of embedding operators is commutative, 
generated by the commuting embedding operators $O\brackx_{\couple}$ of the connected 
diff-classes $O\brackx\in X$.
\eq
[O\brack{x_{1}}_{\couple},\, O\brack{x_{2}}_{\couple}] =0
\en
The embedding operators are nilpotent in the sense that
\eq
O\brackx^{n}_{\couple} \delta_{M'} = 0 \qquad n\ge n_{0}(O,M')\ge 0
\en
The $O\brack{x}_{\couple}$ are destruction operators.  They decrease the number of 
connected components of the boundary.

From the gluing formula (\ref{eq:invgluingformula})
\eq
\label{eq:invariantgluingformula}
e^{\bq{\cdot} Q}\, D\phi = e^{\bq{\cdot} A^{t}}\, A_{*} 
e^{\int_{X}\delta_{x} \bw dx}
= A_{*} \left(
e^{\int_{X} O_{\couple}\sbrack{x} \bq\bw dx  }
e^{\int_{X}\delta_{x} \bw dx}
\right)
\en
The expression in parentheses is a measure on $\cM'$,
so this equation will express the conditional measure $D\phi_{|\bq}$ on $\GammaY$
as a sum of globally invariant amplitudes.
Evaluating the expression in parentheses is
a problem in the structure of manifolds,
completely independent of the local amplitudes.

The algebras of real-valued functions on $\cX$, $\cX_{0}$ and $\cX'$
are the direct products
\eqg
\Reals_{\cX} =  \mathop\oplus_{x\in\cX}\Reals
\qquad
\Reals_{\cX_{0}} =  \mathop\oplus_{x_{0}\in\cX_{0}}\Reals
\qquad
\Reals_{\cX'} = \Reals_{\cX} \times \Reals_{\cX_{0}} =  \mathop\oplus_{x'\in\cX'}\Reals
\eng
An identity is derived in Appendix~\ref{app:gluingidentity}:
\eqg
\label{eq:gluingidentitybody}
O\brackx_{\couple}  \; e^{\int_{\cX'}   \delta_{x'}\bp'  dx'} 
=
\bv'\brackx(\bp)\partial_{\bp'}
\; e^{\int_{\cX'}  \delta_{x'}\bp'  dx'} 
\\[2ex]
\bp'\in \Reals_{X'}
\qquad
\bp'=(\bp_{0},\bp)
\qquad
\bp_{0}\in \Reals_{X_{0}}
\qquad
\bp\in \Reals_{X}
\eng
where $\bv'\brackx(\bp)\partial_{\bp'}$ is the vector field on 
$\Reals_{\cX'}$
\eq
\label{eq:Ovidentification}
\bv'\brackx =\bv'\brackx(\bp)\partial_{\bp'} =
\int_{\cX'} \sum_{\bm} 
\frac{\bp{}^{\bm}}{\bm!}
\frac{\partial}{\partial\bp'\sbrack{x'}} 
\left(O\brackx_{\couple}  \delta_{M^{\bm}}\right){}_{/X'}
\en
$\left(O\brackx_{\couple}  \delta_{M^{\bm}}\right){}_{/X'}$
is the restriction to $X'$ of the measure $O\brackx_{\couple}  
\delta_{M^{\bm}}$ on $\cM'$.
Equation (\ref{eq:Ovidentification}) identifies the embedding operator acting on $\meas(\cM')$ with
the vector field on $\Reals_{X'}$
via the linear transform
\eq
\delta_{\bp'} \quad\mapsto\quad e^{\int_{\cX'}  \delta_{x'} \bp' dx'} 
\qquad
\meas(\Reals_{X'}) \rightarrow \meas(\cM')
\en
which identifies the derivatives of $\delta_{0}$ with the finite measures
on $\cM'$.
There is a vector field $\bv'\brackx$ for every $x\in 
X$.  They commute with each other.
\eq
[\bv'\brack{x_{1}},\,\bv'\brack{x_{2}}] = 0
\en
$\bv'\brackx(\bp)\partial_{\bp'}$ depends only on $\bp$, independent of $\bp_{0}$,
because $\Ox_{\couple}$ commutes with multiplication by 
$\delta_{M_{0}}$ for every closed diff-class $M_{0}$.

Exponentiate (\ref{eq:gluingidentitybody})
\eqg
e^{\int_{X} \Ox_{\couple} \btau\sbrack{x} dx}
\;e^{\int_{\cX'}   \delta_{x'}\bp'_{i}  dx'} 
=
e^{\int_{\cX'}\delta_{x'} \bp'(\btau, \bp'_{i})   dx'} 
\qquad
\btau \in \Reals_{X}
\\[2ex]
\frac{\partial}{\partial \btau\sbrack{x}} \bp' = \bv'\brackx(\bp)
\qquad
\bp'(\btau,\bp'_{i})_{/\btau=0} =  \bp'_{i}
\eng
$\btau \mapsto \bp'(\btau, \bp'_{i}) $ is the flow of the 
commuting vector fields with initial condition $\bp'=\bp'_{i}$ at $\btau=0$.
The vector fields are independent of $\bp_{0}$ so we first integrate
\eq
\frac{\partial}{\partial \btau\sbrack{x}} \bp = \bv\brackx(\bp)
\qquad
\bp(\btau,\bp_{i})_{/\btau=0} =  \bp_{i}
\en
to get the flow on $\Reals_{X}$,
then integrate
\eqg
\frac{\partial}{\partial \btau\sbrack{x}} \bp_{0} = \bv_{0}\brackx(\bp)
\qquad
\bp_{0}(\btau,\bp'_{i})_{/\btau=0} =  \bp_{0i}
\eng
to get the flow on $\Reals_{X_{0}}$.

For the expression in parenthesis in (\ref{eq:invariantgluingformula})
substitute
\eq
\btau  = \bq \bw
\qquad
\bp'_{i} = (\bp_{0i},\bp_{i})
\quad
\bp_{0i}=0 \quad \bp_{i} = \bw
\en
to get
\eq
e^{\int_{X} O_{\couple}\sbrack{x} \bq\bw dx  }
e^{\int_{X}\delta_{x} \bw dx}
=
e^{\int_{\cX'}\delta_{x'} \bz'(\bq, \bw)   \bw' dx'} 
\en
where $\bz'(\bq, \bw)$ is the reparametrized flow
\eq
\bz'(\bq, \bw)\,\bw'  = \bp'(\btau,\bp'_{i}) = \bp'(\bq \bw,(0,\bw)) 
\en
Now substitute in  (\ref{eq:invariantgluingformula}).
\eqa
e^{\bq{\cdot} Q}\, D\phi &= A_{*} \left(
e^{\int_{X} O_{\couple}\sbrack{x} \bq\bw dx  }
e^{\int_{X}\delta_{x} \bw dx}
\right)
\\[2ex]
&= A_{*} \left(e^{\int_{\cX'}\delta_{x'} \bz'(\bq, \bw)   \bw' dx'} 
\right)
\\[2ex]
&= e^{\int_{\cX'}A' \bz'(\bq, \bw)   \bw' dx'} 
\\[2ex]
&= e^{\int_{\cX_{0}}A_{0} \bz_{0}(\bq, \bw)   \bw_{0} dx_{0}}
e^{\int_{\cX}A \bz(\bq, \bw)   \bw dx}
\ena
The $A_{0}\brack{x_{0}}$ are numbers ---
amplitudes of closed diff-classes.  
So the expectation value is
\eq
\expval{e^{\int_{\cX} \bq \bQ \,\bw dx}}
= e^{\int_{\cX_{0}}  \bz_{0}(\bw,\bq) A_{0} \bw_{0} dx_{0}}
\en
and the conditional probability measure is
\eq
D\phi_{|\bq} = 
\frac{e^{\int_{\cX} \bq \bQ \,\bw dx} D\phi}
{\expval{e^{\int_{\cX} \bq \bQ \,\bw dx}}}
=e^{\int_{\cX}  \bz(\bw,\bq) A \bw dx}
\en
The conditional probability measure is given by a sum of amplitudes
over the ensemble $\cM$
with a modified ``bulk action''
expressing the embedded quiescent experiments
as ``bulk operators''
or ``bulk sources''.

\section{Embedding principle}

The \emph{embedding principle} is
that the function $\bz'(\bw,\bq)$ should 
not depend on $\bw$.
Change variables in (\ref{eq:gluingidentitybody}),
parametrizing the flow by $\bq$ in place of $\btau$
and parametrizing $\Reals_{X'}$ by $\bz'$ in place of $\bp'=\bz'\bw'$.
\eqg
\bw\brackx O\brackx_{\couple}  \; e^{\int_{\cX'}   \delta_{x'}\bz' \bw'  dx'} 
=
\bu'\brackx(\bz,\bw')\partial_{\bz'}
\; e^{\int_{\cX'}  \delta_{x'}\bz' \bw'  dx'} 
\eng
\eq
\bu'\brackx(\bz,\bw')\partial_{\bz'} =
\int_{\cX'} \sum_{\bm} 
\frac{\bz{}^{\bm}}{\bm!}
\left(\frac{\bw\brackx\bw^{\bm} }{\bw'\brack{x'}}\right)
\frac{\partial}{\partial\bz'\sbrack{x'}} 
\left(O\brackx_{\couple}  \delta_{M^{\bm}}\right){}_{/X'}
\en
All dependence on $\bw$ is collected in the number
\eq
\frac{\bw\brackx\bw^{\bm} }{\bw'\brack{x'}}
\en
$\bz'(\bw,\bq)$ is independent of $\bw$ if
\eq
\label{eq:embed1}
\bw'\brack{x'} = \bw\brackx  \bw^{\bm} 
\en
whenever the connected diff-class $\Ox$ can be embedded in the interior of
the connected diff-class $M'\brack{x'}$ with 
complement the diff-class $M^{\bm}$.  
The weights $\bw'\brack{x'} =1$ trivially satisfy the embedding condition
(\ref{eq:embed1}).

With embedding condition (\ref{eq:embed1}),
the quiescent expectation value is
\eq
\label{eq:expval}
\expval{e^{\int_{\cX} \bq \bQ \,\bw dx}}
= e^{\int_{\cX_{0}}  \bz_{0}(\bq) A_{0} \bw_{0} dx_{0}}
\en
and the conditional measure is
\eq
\label{eq:condprob}
D\phi_{|\bq} = 
\frac{e^{\int_{\cX} \bq \bQ \,\bw dx} D\phi}
{\expval{e^{\int_{\cX} \bq \bQ \,\bw dx}}}
=e^{\int_{\cX}  \bz(\bq) A \bw dx}
\en
The \emph{embedding functions} $\bz_{0}(\bq),\,\bz(\bq)$
are the flow in $\bq$ of the commuting vector fields 
$\bv'\brackx(\bp)\partial_{\bp'}$
starting from the initial point $\bp'_{i} = (0,1)$.
\eq
\bz_{0}(\bq) = \bp_{0}(\bq,(0,1))
\qquad
\bz(\bq) = \bp(\bq,(0,1))
\en

The quiescent experiments parametrized by $\bq$ are represented
in (\ref{eq:expval}) and (\ref{eq:condprob}) as ``bulk operators''
parametrized by the embedding functions $\bz_{0}(\bq)$ and $\bz(\bq)$.
These bulk observables are universal in form, independent of the local EQG.
Non-quiescent local experiments can also be represented as bulk observables
in essentially the same way, although the technology is more 
complicated because the gluing axiom will not apply.

Condition (\ref{eq:embed1}) on the weights has several equivalent 
forms.
First, it is equivalence to invariance of the weights
under the embedding map $E'(\hat O,\hat M)$
\eq
\label{eq:embed4}
E'_{*} ( w dO, w dM)  = \bw' dx'
\en
$E'(\hat O,\hat M)$ is the diff-class formed by attaching the 
relative diff-class $\hat O$ at its boundary to
the relatively diff-class $\hat M$,
when $\partial \hat O$ is a submanifold of $\partial \hat M$.

Define the \emph{reduced weight} of the general diff-class $M'^{\bm'} 
\in \cM'$
\eq
\wred(M'^{\bm'})=\bw'^{\bm'}
\qquad
M'^{\bm'}= \sqcup_{x'} M'\brack{x'}^{\bm'\sbrack{x'}}
\en
Equation (\ref{eq:embed1}) becomes
\eq
\wred(M'\brack{x'}) = \wred(O\brack{x}) \wred(M)
\en
which is equivalent to
\eq
\wred(M') = \wred(O\brack{x}) \wred(M'-O\brack{x})
\en
whenever $O\brack{x}$ can be embedded in $M'$ with complement
$M'-O\brack{x}$, for any diff-class $M'$.
The equivalence follows from the fact that
the connected $O\brackx$ is necessarily embedded in one of the connected 
components of $M'$.
If $O\in \cM$ embeds in $\cM'$ it can be done so by a series of 
embeddings of the connected components of $O$, so another equivalent 
form of the embedding condition is
\eq
\label{eq:embed2}
\wred(M') =  \wred(O) \wred(M'-O)
\qquad \forall M'\in \cM', \; O \in \cM
\en
For any diff-class $N\in \cN$,
the cylinder $I\times N$ embeds in $I\times N$ with complement
$\sqcup^{2}(I\times N)$ so
\eq
\wred(I\times N) = \wred(I\times N)^{3}
\qquad
\wred(I\times N) = -1,0, \text{ or }  1
\en
The possibility $\wred(I\times N) =-1$ is presumably eliminated by 
the positivity condition on $D\phi$.
The  possibility $\wred(I\times N) =0$ amounts to truncating
the local EQG by deleting $\Ny$ from $\cY$
and deleting from $\cM'$ all the diff-classes that have $I\times \Ny$ 
embedded,
for some subset of the $\Ny\in Y$.
So we assume $\wred(I\times N) =1$ for all $N$.

Suppose $M''$ is a diff-class whose boundary contains $N\sqcup N$
and $M'$ is a diff-class formed by attaching $I\times N$ to $M''$.
Then
\eq
\label{eq:embed3}
\wred(M') = \wred(I\times N) \wred(M'') = \wred(M'')
\en
Taking $M'' = O \sqcup (M'-O)$
shows that (\ref{eq:embed3})
is equivalent to (\ref{eq:embed2}).
If $\wred(M')$ satisfies the embedding condition (\ref{eq:embed3})
then it can be absorbed into the local amplitude,
replacing $\bA$ with $\wred \bA$.  If $\bA$ 
satisfies the factorization and gluing axioms then so does
$\wred \bA$.
So the solutions of the embedding condition are coupling constants ---
free parameters --- of the local EQG.
They are \emph{global coupling constants}
because they are in every local EQG.
The conditions on the weights provided by the
combination of the connectedness principle and the embedding 
principle are maximally restrictive --- they determine the weights 
up to the global coupling constants.

Still another equivalent form of the embedding condition is in terms of
cobordisms.  Suppose $\tilde C_{3}= \tilde C_{2}\circ \tilde C_{1}$ is a 
composition of cobordisms.  Let $N$ be the diff-class in between --- 
the target of $\tilde C_{1}$ and the source of $\tilde C_{2}$.  Then $C_{3}-I\times N = C_{1}\sqcup C_{2}$ so
\eq
\label{eq:embed5}
\wred(C_{3}) = \wred(C_{1})\wred(C_{2})
\qquad \tilde C_{3}= \tilde C_{2}\circ \tilde C_{1}
\en
$\tilde C \mapsto \wred(C)$ is a  character 
on the cobordisms,
taking composition of cobordisms to multiplication of positive real 
numbers,
invariant under diffeomorphisms.

In $d=2$ dimensions, 
\eqa
X &= X_{+}\sqcup X_{-}
\\[2ex]
X_{+} &= \{ M_{+}[g,n],\; g\ge 0, \,n\ge 0\}
\\
&= \{\text{orientable surfaces of genus $g$ with $n$ boundary components}\}
\\[2ex]
X_{-} &= \{ M_{-}[g,n],\; g\ge 1, \,n\ge 0\}
\\
&= \{\text{unorientable surfaces of genus $g$ with $n$ boundary components}\}
\ena
The solutions of the embedding condition are
\eq
\begin{alignedat}{2}
\wred( M_{+}[g,n])  &= \wred( M_{+}[0,1]) ^{\chi} \qquad& \chi &= 2-2g-n
\\[2ex]
\wred( M_{-}[g,n])  &= \wred( M_{+}[0,1]) ^{\chi}\,
\wred(M_{-}[1,1])^{g} \qquad& \chi &= 2-g-n
\end{alignedat}
\en
$\chi$ is the Euler number.
$M_{+}[0,1]$ is the 2-disk.  $M_{-}[1,1]$ is the Moebius band.
The two weights $\wred( M_{+}[0,1]) $ and $\wred(M_{-}[1,1])$ are the global 
coupling constants.

\section{Questions and comments}

\subsection{Wick rotate to real time QG}

The only evident opportunity for analytic continuation from imaginary time to 
real time is in an individual observer $\Ox$, an individual 
$d$-manifold with non-empty boundary.
If $\Ox$ is a real slice in a diff-class of complex $d$-manifolds
$\Ox_{\Complexes}$
then there could be a Wick rotation --- an analytic continuation
to another real slice.
The observables of the local EQG would analytically continue to the 
observables of a Lorentz-signature local quantum gravity.
Measurements -- correlation functions of Lorentz-signature observables 
--- would be given by analytic continuation from the correlation 
functions of the EQG.
The Hartle-Hawking state on $\Ox$ 
would be the ``state'' of the real time quantum gravity
in the sense that it determines all the real time correlation 
functions.

In principle, real time measurements 
in $\Ox$  could test whether the boundary condition
on $\Ox$ is exactly
the Hartle-Hawking state
or departs from it,
looking for example for evidence of the presence of 
other observers
coupled to the ensemble of $d$-manifolds.

\subsection{Signalling}

The conditional probablity
measure (\ref{eq:condprob}) in the presence of the quiescent observable is
\eq
D\phi_{|\bq} = 
\frac{e^{\int_{\cX} \bq \bQ \,\bw dx} D\phi}
{\expval{e^{\int_{\cX} \bq \bQ \,\bw dx}}}
=e^{\int_{\cX}  \bz(\bq) A \bw dx}
\en
Write $D\phi_{|\bq}$ as the  convolution
\eqg
D\phi_{|\bq} = D\phi
\;
D\Delta \phi
\qquad
D\phi = e^{\int_{\cX}   A \bw dx}
\qquad
D\Delta \phi = 
e^{\int_{\cX}  [\bz(\bq)-1] A \bw dx}
\eng
Let $\phi$ be the random variable associated to the probability 
measure $D\phi$,
let $\phi_{|\bq}$ be the random variable associated to $D\phi_{|\bq}$,
and let $\Delta\phi$ be the random variable associated to 
$D\Delta\phi$.  The covolution of the probability measures
is equivalent to the equation
between random variables
\eq
\phi_{|\bq} = \phi +\Delta\phi
\qquad
\Delta\phi = \phi_{|\bq} - \phi
\en
The probability measure $D\Delta\phi$ with random variable 
$\Delta\phi$ is
called the \emph{distribution of differences}.
At $\bq=0$,   $D\Delta\phi = \delta_{0}$,
 $\Delta\phi =0$.  As $\bq$ varies from 0, the 
distribution $D\Delta\phi$ broadens
from the delta-function at 0 in $\GammaY$  to a probability measure 
concentrated near $0$.  The random variable $\Delta\phi$ is the 
microscopic signal of the quiescent observable $e^{\bq\cdot\bQ}$ coupled to the ensemble.
The macroscopic random variable $\Delta\bphi = T_{*} \Delta\phi$ is the macroscopic 
signal seen by experiments coupled to the conditional ensemble.

The ensemble is acting as a communication channel.
The observing experiment detects the signal
--- the random variable $\Delta\bphi$ ---
of the presence of the quiescent experiment.
The flow of information is
from the quiescent experiment coupled to the ensemble,
through the ensemble, to the observing experiment coupled to the 
ensemble.
Equivalently, there is a mixed boundary/bulk picture
where the quiescent experiment is embedded in the ensemble.
The ensemble conditional on the embedding
provides the signal to the observing experiment.
And there is the pure bulk description where both
observing experiment and quiescent experiment
are embedded in an ensemble of closed diff-classes.

Any observable source  $\int_{X} \bPhi(x) \bw dx$ could have been used
instead of the quiescent source.
The quiescent observable is singled out because it is mathematically 
natural and because the construction of the bulk
quiescent observable
is simplified by the gluing axiom.
For the quiescent source,
the transmission of information from the embedded experiment 
to the bulk manifold can be calculated entirely in terms of the 
manifolds, without reference to the local EQG.
For a general source, both are involved.

\subsection{Averaging over the mapping class group}

Statistical EQG in $d\ge 3$ dimensions needs
averaging measures $\mu dg$ on the infinite mapping class groups $G$
of the closed \dmo-manifolds.
Averaging over the action of $G$ in a finite dimensional 
representation should project on the subspace of $G$-invariant vectors.
There should be a topology on $G$ weaker than the discrete topology,
a topology in which the matrix elements of finite dimensional 
representations are continuous functions on $G$.
There should be a completion $\bar G$,
adding a set of ``ends''.
The averaging measure would have to be supported on the set of ends,
since the measure of any finite subset of $G$ would have to be zero.

The set of diff-classes $\cM'$ needs a compatible completion 
$\bar\cM'$
on which $G$-averaged quantities such as the measure used in the 
embedding formulas,
\eq
\rho(x,M,x')dx' = \int_{\MCG(\partial\hat M)}
\delta_{ E'(\hat O\brackx, g \hat M)}\left(x' \right ) dx'
\en
will be well-defined.

The formulas
\eq
D\phi =e^{\int_{\cX}   A \bw dx}
\qquad
D\phi_{|\bq} =e^{\int_{\cX}  \bz(\bq) A \bw dx}
\en
suggest there might be a notion of locality in $X$
in analogy with classical 
statistical field theory.
Such a notion of locality might be implicated
in the completion of $X$.

\subsection{Meaning of the embedding principle}
The embedding principle --- that  $\bz'(\bw,\bq)$ 
is independent of $\bw$ ---
is mathematically natural and it is expressed in terms of the 
correlation functions of statistical EQG ---
in terms of the expectation value and
the conditional measure of the quiescent observable ---
but it lacks physical meaning.
The principle is purely technical.
It amounts to requiring $\bz'(\bw,\bq) = \bz'(1,\bq)$,
which builds in from the start that $\bw=1$ is allowed.
This is a natural requirement,
but perhaps not quite in the spirit of the project to determine the weights
on principle.

A connection
of the embedding principle
to ``bulk locality'' is suggested by
the embedding condition in the form of
equation (\ref{eq:embed4})
which says that the measure on diff-classes is produced from the measures on their 
embedded parts.
Equations
(\ref{eq:expval})
and (\ref{eq:condprob})
suggest
that a locality principle might be involved
in the form of the ``bulk action''  as an integral over $X'$.

\subsection{Form of the constraints on the weights}

The constraints on the weights that they should produce a globally 
invariant Hartle-Hawking state
expresses the principle that every observer should be in a locally 
invariant state.
There is a gap in the argument.
Global invariance implies 
local invariance on every 
$d$-manifold with boundary.
But the converse is not established.
Similarly, the connectedness principle in the microscopic system
implies that all the connected correlation functions of observers
are given by summing over the connected diff-classes of $d$-manifolds,
but the converse is not established.
There is the same gap in the embedding principle.
It seems a reasonable assumption that 
the properties of the microscopic system
are needed to get the desired macroscopic properties
for all possible macroscopic experiments.

\subsection{Re-formulate the gluing axiom}

The gluing axiom as expressed in Section \ref{sect:GluingAxiom}
is awkward.
The amplitude of a local EQG is completely encoded in
the locally invariant states
$\bA'\brack{x'} \in \bcS'\brack{x'}$.
The only further constraint on $\bA'$ is the gluing axiom,
which should have a locally invariant formulation.
It should not be necessary to choose representative \dmo-manifolds in 
the boundary diff-classes
in order to express the gluing axiom.

\subsection{Complex EQG}

In complex EQG, the manifolds are oriented and the state spaces are complex.
Diffeomorphisms are orientation preserving.
Reversing orientation is complex conjugation on the states.
In the gluing operation, two \dmo-manifolds are identified
by an \emph{orientation reversing} diffeomorphism.
The quadratic form in the gluing operation is hermitian.
Otherwise everything is the same as in real EQG.

\subsection{Global coupling constants}

The global coupling constants are
the solutions $\bw'[x']$ of the embedding condition 
expressed by any of the equivalent equations
(\ref{eq:embed1}) or (\ref{eq:embed4}) or (\ref{eq:embed2})
or (\ref{eq:embed3}) or (\ref{eq:embed5}).
They form a natural semigroup in each dimension $d$
which undoubtedly is known in mathematics under another name.

\subsection{Finiteness}

It is hard to see how statistical EQG could make sense
unless the state spaces are finite dimensional,
at least in some effective sense.
The amplitude of a torus $S^{1}\times N$ should be the dimension of 
the state space on $N$.
If the amplitude is finite then the state space must be finite 
dimensional.
This might be accomplished by 
making the states spaces graded vector spaces,
adding odd fermionic states to cancel even bosonic states.
Otherwise,
getting around the finiteness requirement would seem to require
some significant elaboration of the structure.

\vskip2ex
\section*{Acknowledgments}
\phantomsection
\addcontentsline{toc}{section}{\numberline{}Acknowledgments}
This work was supported by the New High Energy Theory Center
of Rutgers University.
I am grateful to Anindya Banerjee, Gregory Moore and Shehryar 
Sikander for their 
comments and questions.
I thank Hongbin Sun for
discussions of differential topology in $d=3$ dimensions.

\bibliographystyle{ytphys}
\raggedright
\bibliography{Literature}
\phantomsection
\addcontentsline{toc}{section}{\numberline{}References}

\appendix

\setlength\parindent{2em}

\section{Summary of the global structure}
\label{app:summary}

\vspace*{2ex}

\noindent
$$
\begin{array}{c@{\qquad\quad}c@{\qquad}l}
\multicolumn{3}{l}{\bullet\;\; \text{\bf locally invariant macroscopic system}}
\\[2ex]
&\cX =\big\{\Mx\big\}
&  \text{macroscopic objects/observers} 
\\[2ex]
&\Gamma_{\cX} = \mathop\oplus\limits_{x\in \cX} \bcSx
& \text{macroscopic states/observer states}
\\[3ex]
&\bcS =\Sym(\GammaX) = \meas(\Gamma_{\cX}) = \mathop\oplus\limits_{M\in \cM} \bcS(M)
& \text{measures on the macroscopic states}
\\[3ex]
& D\bphi \in \bcS
& \text{the macroscopic probability measure}
\\[4ex]
\multicolumn{3}{l}{\bullet\;\; \text{\bf globally invariant microscopic system}}
\\[2ex]
&\cY=\big\{\Ny\big\}
& \text{microscopic objects/sensors} 
\\[2ex]
&\Gamma_{\cY} = \mathop\oplus\limits_{y\in \cY} \cSy
& \text{microscopic states/sensor states}
\\[3ex]
&\cS=\Sym(\GammaY) = \meas(\Gamma_{\cY}) = \mathop\oplus\limits_{N\in \cN} \cS(N)
& \text{measures on the microscopic states}
\\[3ex]
& D\phi\in\cS
& \text{the microscopic probability measure}
\\[4ex]
\multicolumn{3}{l}{\bullet\;\; \text{\bf from global to local} }
\\[2ex]
&\GammaY \xrightarrow{T} \GammaX
& \text{microscopic state determines macroscopic state}
\\
&& \text{\quad (globally invariant states are locally invariant)} 
\\[2ex]
& D\phi \mapsto D\bphi = T_{*}D\phi
&  \text{microscopic probabilities determine macroscopic}
\\[3ex]
& \int_{\GammaY} T^{*}\bPhi D\phi = \int_{\GammaX} \bPhi \, D\bphi 
& \text{microscopic expectation values determine}
\\
&& \text{\quad macroscopic correlation functions}
\\[4ex]
\multicolumn{3}{l}{\bullet\;\; \text{\bf from local to global}}
\\[2.5ex]
&\bA \in \GammaX\qquad \bA\brackx \in \bS\brackx
& \text{locally invariant amplitude}
\\[3ex]
&\bcSx \xrightarrow{P} \cS(\partial \Mx)
& \text{projection on the globally invariant states} 
\\[3ex]
& A = P \bA\in \mathop\oplus\limits_{x} \cS(\partial \Mx)\subset \cS
& \text{globally invariant amplitude }
\\[3ex]
& \bw dx
& \text{measure on $\cX$}
\\[3ex]
& wdM = e^{\int_{X}\delta_{x} \bw dx}
& \text{measure on $\cM$}
\\[3ex]
& D\phi= A_{*} wdM = e^{\int_{\cX} A \,\bw dx}
& \text{globally invariant Hartle-Hawking state}
\end{array}
$$

\newpage
\section{Embedding vector fields}
\label{app:gluingidentity}

\subsection{Embedding operators}

For $O\in \cM$,
the \emph{embedding} operator $O_{\couple}$
is the linear operator on measures on $\cM'$
\eqg
O_{\couple}\colon \meas(\cM') \rightarrow \meas(\cM')
\\[2ex]
O_{\couple} \delta_{M'}
=
\left\{
\begin{array}{c@{\qquad}l}
\displaystyle\int_{\MCG(\partial\hat M')}
\delta_{E'(\hat O, g \hat M')}
\,\mu dg
&
\partial O \subseteq \partial M'
\\[3ex]
0 & \partial O \not\subseteq \partial M'
\end{array}
\right .
\eng
$E'(\hat O, g \hat M')$ is the diff-class formed by attaching $O$ to 
$M'$ by
\begin{enumerate}
\item
choosing a representative  $\tN_{M}$ of 
$\partial M$ and a representative $\tN_{O}$ of $\partial O$ such that
$\tN_{M} = \tN_{O} \sqcup \tN'$
for some $\tN'$,
\item then choosing relative diff-classes $\hat O$ and
$\hat M'$ representing $O$ and $M'$,
\item then identifying $\tN_{O} = \partial 
\hat O$ with $\tN_{O} \subset \partial g\hat M'$ to form the diff-class
$E'(\hat O, g \hat M')$ with boundary $N'$.
\end{enumerate}
Averaging over the mapping class group removes all dependence on 
the arbitrary choices.
$O$ is embedded in the interior of $E'$ with complement $M'$.
The embedding operator is multiplicative
\eq
(O_{1}\sqcup O_{2})_{\couple} = O_{1\couple}O_{2\couple} = O_{2\couple} O_{1\couple}
\en
The algebra of embedding operators is commutative, 
generated by the embedding operators of the connected 
$O\brackx$,
\eq
[O\brack{x_{1}}_{\couple},\, O\brack{x_{2}}_{\couple}] =0
\en
The embedding operators are nilpotent in the sense that
\eq
O_{\couple}\brackx^{n} \delta_{M'} = 0 \qquad n\ge n_{0}(O,M')\ge 0
\en
The $O\brack{x}_{\couple}$ are destruction operators.  They decrease the number of 
connected components of the boundary.

\subsection{Transform to embedding vector fields}

$\cM$, $\cM_{0}$ and $\cM'$  are the direct product monoids
\eqg
\NN_{\cX} =  \mathop\oplus_{x\in\cX}\NN x
\qquad
\NN_{\cX_{0}} =  
\mathop\oplus_{x_{0}\in\cX_{0}}\NN x_{0}
\qquad
\NN_{\cX'} = \NN_{\cX} \times \NN_{\cX_{0}} =  
\mathop\oplus_{x'\in\cX'}\NN x'
\eng
The algebras of real-valued functions on $\cX$, $\cX_{0}$ and $\cX'$
are the direct products
\eqg
\Reals_{\cX} =  \mathop\oplus_{x\in\cX}\Reals
\qquad
\Reals_{\cX_{0}} =  \mathop\oplus_{x_{0}\in\cX_{0}}\Reals
\qquad
\Reals_{\cX'} = \Reals_{\cX} \times \Reals_{\cX_{0}} =  \mathop\oplus_{x'\in\cX'}\Reals
\eng
Transform measures on $\Reals_{X'}$ to measures on $\NN_{X'} = \cM'$.
\eq
\delta_{\bp'} \quad\mapsto\quad e^{\int_{X'} \bp'\sbrack{x'} \delta_{x'} dx'}
\qquad
\delta_{0} \mapsto \delta_{0}
\qquad
\bp'=(\bp_{0},\bp)\in \Reals_{X'}
\en
The derivatives of $\delta_{0}$  map to the finitely supported measures on 
$\cM'$.

Consider the measure on $\cM'$
\eq
\label{eq:vmeas}
v\brackx(\bp) =
e^{-\int_{\cX'} \delta_{x'} \bp'  dx'} O\brackx_{\couple}  \, e^{\int_{\cX'}  
\delta_{x'} \bp'  dx'} 
\en
$v\brackx(\bp)$ is independent of $\bp_{0}$ because 
$O\brackx_{\couple}$ commutes with multiplication by $\delta_{M_{0}}$ 
for every closed diff-class $M_{0}$.
$v\brackx(\bp)$ is supported on the set of connected diff-classes 
$\cX'\subset \cM'$.
\eq
v\brackx(\bp) = v\brackx(\bp,x')dx'
= \int_{\cX'} \delta_{x'}\,v\brackx(\bp,x')\,  dx'
\en
Therefore
\eqa
\label{eq:FTOcouple}
O\brackx_{\couple}  \, e^{\int_{\cX'}   \delta_{x'} \bp'dx'} 
&=
v\brackx(\bp)\, e^{\int_{\cX'}   \delta_{x'} \bp'dx'} 
\\[2ex]
&=
\int_{\cX'}  \delta_{x'}\,v\brackx(\bp,x')\, dx'\, e^{\int_{\cX'}  \delta_{x'}  \bp'dx'} 
\\[2ex]
&=
\int_{\cX'}\frac{\partial}{\partial_{\bp'\sbrack{x'}}} v\brackx(\bp,x')\,  dx'\, e^{\int_{\cX'}  \bp' \delta_{x'} dx'} 
\ena
which is
\eq
\label{eq:Ovequiv}
O\brackx_{\couple}  \, e^{\int_{\cX'} \delta_{x'} \bp'  dx'} 
=
\bv'\brackx(\bp)\partial_{\bp'}
\, e^{\int_{\cX'} \delta_{x'}  \bp' dx'} 
\en
where $\bv'\brackx$ is the vector field on 
$\Reals_{\cX'}$
\eq
\bv'\brackx=\bv'\brackx(\bp)\partial_{\bp'} =
\int_{\cX'} \frac{\partial}{\partial_{\bp'\sbrack{x'}}} v\brackx(\bp,x')\, dx'
\en
$O\brackx_{\couple}$ acting on $\meas(\cM')$
is the transform
of the vector field $\bv'\brackx(\bp)\partial_{\bp'}$
acting on $\meas(\Reals_{\cX'})$.
Since $v\brackx(\bp)$ is supported on $\cX'$,  equation (\ref{eq:vmeas})
can be written
\eq
v\brackx(\bp) = O\brackx_{\couple}  \, e^{\int_{\cX}  \bp \delta_{x}
dx} 
{}_{/X'}
=
\sum_{\bm} \frac{\bp{}^{\bm}}{\bm!}
\left(O\brackx_{\couple}  \delta_{M^{\bm}}\right){}_{/X'}
\en
The vector field is
\eq
\bv'\brackx =\bv'\brackx(\bp)\partial_{\bp'} =
\int_{\cX'} \sum_{\bm} 
\frac{\bp{}^{\bm}}{\bm!}
\frac{\partial}{\partial\bp'\sbrack{x'}} 
\left(O\brackx_{\couple}  \delta_{M^{\bm}}\right){}_{/X'}
\en

\subsection{Embedding flow}

Exponentiate equation (\ref{eq:Ovequiv}).
\eq
e^{\int_{X} \btau\sbrack{x} \Ox_{\couple} dx}
\;e^{\int_{\cX'}  \bp'_{i} \delta_{x'}  dx'} 
=
e^{\int_{\cX'} \bp'(\btau, \bp'_{i}) \delta_{x'}  dx'} 
\qquad
\btau\brackx \in \Reals
\en
where $\btau \mapsto \bp'(\btau, \bp'_{i}) $ is the flow of the 
commuting vector fields.
\eq
\frac{\partial}{\partial \btau\sbrack{x}} \bp' = \bv'\brackx(\bp)
\qquad
\bp'(\tau,\bp'_{i})_{/\tau=0} =  \bp'_{i}
\en
The vector fields are independent of $\bp_{0}$ so we first
integrate the flow on $\Reals_{X}$,
\eq
\frac{\partial}{\partial \btau\sbrack{x}} \bp = \bv\brackx(\bp)
\qquad
\bp(\tau,\bp_{i})_{/\tau=0}  =  \bp_{i}
\en
then derive the flow on $\Reals_{X_{0}}$ from the flow on 
$\Reals_{X}$.
\eqg
\frac{\partial}{\partial \btau\sbrack{x}} \bp_{0} = \bv_{0}\brackx(\bp)
\qquad
\bp_{0}(\tau,\bp'_{i})_{/\tau=0}  =  \bp_{0i}
\\[2ex]
\bp_{0}(\tau,\bp'_{i}) = \bp_{0i}+  \int_{0}^{\btau}
\int_{X}
\bv_{0}\brackx(\bp(\btau',\bp_{i}))
d\btau'\brackx 
\;dx
\eng

\end{document}